\documentclass[aps, prd, onecolumn, tightenlines, notitlepage, superscriptaddress, nofootinbib, preprintnumbers, floatfix,showkeys,11pt]{revtex4-2}

\usepackage[normalem]{ulem}
\usepackage{amstext}
\usepackage{amssymb}
\usepackage{amsmath}
\usepackage{graphicx}
\graphicspath{{plots/}}
\usepackage{url}
\usepackage{color}
\usepackage{ulem}
\usepackage[utf8]{inputenc}
\pdfoutput=1
\usepackage{textcomp}
\usepackage{booktabs,siunitx,array,threeparttable}
\usepackage{epsfig,amsfonts,mathrsfs,amsmath,amssymb,graphicx,color,slashed,multirow}

\usepackage[x11names]{xcolor}
\usepackage[colorlinks]{hyperref}

\usepackage{textgreek} 
\usepackage{caption, subcaption} 
\usepackage{booktabs} 

\definecolor{vdrgreen}{rgb}{0.0, 0.7, 0.0}

\definecolor{lightapricot}{rgb}{0.99, 0.84, 0.69}
\definecolor{nicered}{rgb}{0.7,0.1,0.1}
\definecolor{nicegreen}{rgb}{0.1,0.5,0.1}
\definecolor{coral}{rgb}{1.0, 0.5, 0.31}
\definecolor{blue(ncs)}{rgb}{0.0, 0.53, 0.74}
\definecolor{darkspringgreen}{rgb}{0.09, 0.45, 0.27}
\definecolor{seagreen}{rgb}{0.18, 0.55, 0.34}

\definecolor{cadmiumgreen}{rgb}{0.0, 0.42, 0.24}
\definecolor{chromeyellow}{rgb}{1.0, 0.65, 0.0}
\definecolor{darkturquoise}{rgb}{0.0, 0.81, 0.82}
\definecolor{denim}{rgb}{0.08, 0.38, 0.74}
\definecolor{purple(x11)}{rgb}{0.63, 0.36, 0.94}
\definecolor{red(ncs)}{rgb}{0.77, 0.01, 0.2}
\definecolor{ruddypink}{rgb}{0.88, 0.56, 0.59}
\definecolor{slateblue}{rgb}{0.42, 0.35, 0.8}
\definecolor{airforceblue}{rgb}{0.36, 0.54, 0.66}
\definecolor{orange(colorwheel)}{rgb}{1.0, 0.5, 0.0}

\makeatletter
    \newcommand{\colorboxed}[3][white]{\fcolorbox{#2}{#1}{\m@th$\displaystyle#3$}}
\makeatother

\AtBeginDocument{\hypersetup{citecolor=seagreen,linkcolor=seagreen,urlcolor=seagreen}}
\usepackage{appendix}

\begin{document}

\title{{\LARGE Neutrino oscillation bounds on  quantum decoherence}}
\author{Valentina De Romeri}
\email{deromeri@ific.uv.es}
\affiliation{Instituto de F\'{i}sica Corpuscular (CSIC-Universitat de Val\`{e}ncia), Parc Cient\'ific UV C/ Catedr\'atico Jos\'e Beltr\'an, 2 E-46980 Paterna (Valencia) - Spain}

\author{Carlo Giunti}
\email{carlo.giunti@to.infn.it}
\affiliation{Istituto Nazionale di Fisica Nucleare (INFN), Sezione di Torino, Via P. Giuria 1, I--10125 Torino, Italy}

\author{Thomas Stuttard}
\email{thomas.stuttard@nbi.ku.dk}
\affiliation{Niels Bohr Institute, University of Copenhagen, DK-2100 Copenhagen, Denmark}

\author{Christoph A. Ternes}
\email{ternes@to.infn.it}
\affiliation{Istituto Nazionale di Fisica Nucleare (INFN), Sezione di Torino, Via P. Giuria 1, I--10125 Torino, Italy}
\affiliation{Dipartimento di Fisica, Universit\`a di Torino, via P. Giuria 1, I--10125 Torino, Italy}


\begin{abstract}
We consider quantum-decoherence effects in neutrino oscillation data. Working in the open quantum system framework we adopt a phenomenological approach that allows to parameterize the energy dependence of the decoherence effects. We consider several phenomenological models. We analyze data from the reactor experiments RENO, Daya Bay and KamLAND and from the accelerator experiments NOvA, MINOS/MINOS+ and T2K. We obtain updated constraints on the decoherence parameters quantifying the strength of damping effects, which can be as low as $\Gamma_{ij} \lesssim 8 \times 10^{-27}$ GeV at 90\% confidence level in some cases. We also present sensitivities for the future facilities DUNE and JUNO.
\end{abstract}
\maketitle

\section{Introduction}

The outstanding experimental progress of the past years has allowed to establish a clear picture of neutrino oscillations~\cite{Kajita:2016cak,McDonald:2016ixn} and to measure most neutrino oscillation parameters with unprecedented accuracy~\cite{deSalas:2020pgw,Esteban:2020cvm,Capozzi:2021fjo}. While few open questions remain to be answered, such as the determination of possible leptonic CP violation, the neutrino mass ordering and the octant of the atmospheric angle, the next generation of neutrino experiments will likely allow to close up the neutrino physics picture and to determine all oscillation parameters with good precision.

The established oscillation paradigm, and the consequent interpretation of current and forthcoming neutrino oscillation data, relies on the three-neutrino mass and mixing scheme. According to this scenario,
neutrinos propagate as a superposition of mass and flavor eigenstates. During propagation, each component mass eigenstate evolves with a different frequency thus resulting in the phenomenon of flavor conversion.
Since neutrinos interact only very weakly with other matter particles and since they are stable, this quantum superposition effect is maintained over macroscopic distances. In general, the study of neutrino oscillations considers the neutrino to be isolated from its environment, and the oscillation effects to be coherent. However, neutrino eigenstates may loose their quantum superposition, for example if the neutrino system interacts with a stochastic environment, thus resulting in a loss of coherence and in the damping of neutrino oscillation probabilities -- a phenomenon known as neutrino decoherence.

Different physical origins can lead to similar, nonstandard decoherence effects, see for instance Refs.~\cite{Kiers:1995zj,Ohlsson:2000mj,Beuthe:2001rc,Beuthe:2002ej,Giunti:2003ax,Blennow:2005yk,Farzan:2008eg,Kayser:2010pr,Jones:2014sfa,Akhmedov:2019iyt,Grimus:2019hlq,Naumov:2020yyv,Akhmedov:2022bjs,Krueger:2023skk}. Let us note that the possible loss of flavor-coherence of the neutrino beam can also occur during propagation in standard quantum mechanics, depending on the neutrino wave-packet widths and separation~\cite{Giunti:1991sx,Beuthe:2002ej,Kayser:2010pr,Naumov:2013uia,deGouvea:2020hfl,deGouvea:2021uvg}. We will not consider decoherence from wave-packet separation in this paper.

We will work in the framework of open quantum systems~\cite{Breuer:2002pc}, commonly used in quantum decoherence studies~\cite{Gago:2000qc,Benatti:2000ph,Lisi:2000zt,Benatti:2001fa, Morgan:2004vv,Anchordoqui:2005gj,Fogli:2007tx,Farzan:2008zv, Oliveira:2013nua, Oliveira:2016asf, Jones:2014sfa,  BalieiroGomes:2016ykp, Coelho:2017zes, Coelho:2017byq, Carpio:2017nui,   Coloma:2018idr, Carpio:2018gum, Carrasco:2018sca, Buoninfante:2020iyr, Gomes:2020muc, Ohlsson:2020gxx,Stuttard:2020qfv,Stuttard:2021uyw,Banerjee:2022slh}. We will assume that the oscillation phenomenon can be changed by the interaction of the neutrino subsystem with the environment. 
The fluctuating nature of space-time in a quantum theory of gravity is a commonly cited potential source of a stochastic background that might produce neutrino decoherence effects~\cite{Hawking:1976ra,Ellis:1983jz,Giddings:1988cx,Addazi:2021xuf}.
In this work, however, we will adopt a phenomenological approach and remain agnostic about its origin. By analyzing several neutrino oscillation experiments, we will set stringent limits on possible decoherence effects.

Experimental searches for neutrino quantum decoherence effects have been performed with a vast range of facilities, and assuming different neutrino sources.  In general, the loss of coherence does not prevent neutrino flavor conversion, but modifies the oscillation phenomenon through distance- and energy-dependent damping terms, as the neutrinos propagate away from the source. This suggests that long-baseline experiments are better suited for probing neutrino quantum decoherence. 
Regarding neutrino sources, astrophysical neutrinos are therefore excellent candidates. Solar neutrinos have been used to place the strongest limits on energy-independent quantum decoherence, as shown in \cite{Fogli:2007tx,deHolanda:2019tuf}, in combination with data from the Kamioka Liquid Scintillator Antineutrino Detector (KamLAND). It should be noted, though, that depending on the model of decoherence, all sensitivity to measure decoherence parameters might be lost if neutrino oscillations become averaged, as in the case of solar neutrinos (or the diffuse high energy extragalactic neutrino flux). We will comment on this important point later on.
Earlier works made use of atmospheric neutrinos using Super-Kamiokande data to obtain bounds on the quantum decoherence parameters~\cite{Lisi:2000zt}. Currently, the most stringent limits on decoherence parameters with positive energy dependence ($\Gamma\propto E^{n}, n>0$) have been obtained using atmospheric neutrinos observed at the IceCube Neutrino Observatory~\cite{Coloma:2018idr}. Note that if the origin of the decoherence effects is related to quantum gravity, one would naturally expect $n>0$.
Long-baseline accelerator and reactor experiments, like the Main Injector Neutrino Oscillation Search (MINOS) and KamLAND have also been proven capable of setting bounds on decoherence parameters~\cite{deOliveira:2013dia,BalieiroGomes:2016ykp}. Further analyses include long-baseline accelerator data from MINOS and Tokai to Kamioka (T2K)~\cite{Gomes:2020muc}, as well as
from NuMI Off-Axis $\nu_e$ Appearance (NOvA) and T2K~\cite{Coelho:2017zes}. Sensitivity studies have been presented for the Deep Underground Neutrino Experiment (DUNE) in~\cite{BalieiroGomes:2018gtd,Carpio:2018gum} and for the Jiangmen Underground Neutrino Observatory (JUNO)~\cite{JUNO:2021ydg}.

Building upon these earlier works, we work in the open quantum system framework and analyze data from multiple experiments in order to set up-do-date stringent constraints on the decoherence parameters. Keeping a phenomenological approach, we will parameterize the decoherence effects and their energy dependence in a model-independent way, so to allow for both $n\geq 0$ and $n < 0$. Since there are many possible sources of neutrino decoherence, it is desirable to search for these effects in as many experiments and energy regimes as possible. In this work we compare data from a range of reactor and accelerator experiments to maximize the sensitivity across a broad range of parameter space. As we will show, depending on the energy dependence and on the distance travelled by neutrinos from the source to the detector, different facilities will be more sensitive to the damping effects than others. We analyze data from reactor experiments such as the Reactor Experiment for Neutrino Oscillation (RENO)~\cite{jonghee_yoo_2020_4123573} and Daya Bay~\cite{DayaBay:2018yms}, which allow for high-resolution, high-statistics measurements of the antineutrino flux and consequently of possible damping features. We also consider KamLAND~\cite{Gando:2010aa,kamland_web}, which observes neutrinos from a large number
of nuclear reactor sites with longer baselines, thus providing complementary information. We further analyze data from accelerator experiments, which are sensitive to higher-energy neutrinos. We include the MINOS~\cite{Adamson:2017uda}, NOvA~\cite{NOvA:2018gge,NOvA:2021nfi} and T2K~\cite{Abe:2021gky} experiments. The analysis of reactor and accelerator data yields sensitivity to decoherence effects across the MeV to GeV regimes. Finally, we provide sensitivity analyses for two future facilities:  JUNO~\cite{JUNO:2015zny,JUNO:2021vlw}, a reactor experiment sensitive to the solar oscillation parameters and the atmospheric mass splitting, and especially designed for measuring the neutrino mass ordering, and for the DUNE long-baseline neutrino experiment~\cite{DUNE:2020lwj,DUNE:2020ypp,DUNE:2020mra,DUNE:2020txw}, also expected to determine the neutrino mass ordering, among other goals in its extended physics program.

This paper is organized as follows. We introduce in Section~\ref{sec:theory} the open quantum system formalism and the phenomenological scenarios that we analyze. In Section~\ref{sec:experiments} we present technical details of the experiments and of our statistical analyses. We discuss the results of our analyses in Section~\ref{sec:results}, presenting updated bounds on the decoherence parameters and sensitivities for near-future experiments. We finally draw our conclusions in Section~\ref{sec:concl}.

\section{Theoretical framework}
\label{sec:theory}
In this section we present the theoretical framework that we use to obtain the neutrino oscillation probabilities in the presence of decoherence effects. We adopt the open quantum system approach, treating propagating neutrinos as a subsystem which interacts very weakly with a larger, unknown environment~\cite{Alicki:1105909, Breuer:2002pc}. The interaction of the neutrino system with the quantum environment is the source of damping effects in the oscillations.

The evolution in time of neutrinos in the quantum (sub-)system can be described through the Lindblad  equation~\cite{Lindblad:1975ef,Gorini:1975nb}

\begin{equation}
	\frac{\partial\rho_\nu(t)}{\partial t}=-i[H,\rho_\nu(t)] + \mathcal{D}[\rho_\nu(t)]\,,
    \label{eq:rho_time_dep}
\end{equation}
where $\rho_\nu(t)$ is a Hermitian density matrix describing the neutrino states in the mass basis and $H$ is the Hamiltonian of the neutrino subsystem, which encodes both the standard oscillation effects resulting from the unequal neutrino masses, and any conventional matter effects.
According to the density matrix formalism, the density matrix of mixed quantum states is $\rho_\nu = \sum_j \mathcal{P}_j |\psi_j \rangle \langle \psi_j|$, for a system of $j$ states of probability $\mathcal{P}_j$.
The dissipative term $\mathcal{D}[\rho_\nu(t)]$ encodes the decoherence effects, such that when $\mathcal{D}[\rho_\nu(t)] \to 0$ the standard neutrino oscillations without damping effects are recovered.  It is also a Hermitian matrix which can be expressed as~\cite{Gago:2002na,Benatti:2000ph,Lisi:2000zt,Stuttard:2020qfv}

\begin{equation}
\label{eq:DecTerm}
\mathcal{D}[\rho_\nu(t)]=\frac{1}{2}\sum_{j=1}^{N^2-1} \left([\mathcal{O}_j,\rho_\nu(t) \mathcal{O}_j^{\dagger}]+[\mathcal{O}_j\rho_\nu(t), \mathcal{O}_j^{\dagger}] \right) \, ,
\end{equation}

where $N$ refers to the dimension of the $SU(N)$ Hilbert space defining the neutrino subsystem ($N =3$ when assuming three neutrino flavors) and $\mathcal{O}_j$ are dissipative operators described as $N \times N$ complex matrices
which characterize the coupling of the neutrino subsystem to its environment ($j_\text{max} = 8$ for three neutrino flavors), for example in quantum gravity space-time~\cite{Barenboim:2004wu,Barenboim:2006xt,Benatti:2001fa,Mavromatos:2006yn}. The operators  $\mathcal{O}_j$ act only on neutrino states in such a way that $\sum_j \mathcal{O}^\dagger_j \mathcal{O}_j = \mathbb{I}$ and the trace of $\rho_\nu(t)$ is unchanged. Let us notice that, while being Hermitian, the matrix $\mathcal{D}[\rho_\nu(t)]$ can be nonunitary due to the interactions of the neutrino subsystem with the environment including possible neutrino loss.
The general expression of $\mathcal{D}[\rho_\nu(t)]$ in Eq.~\eqref{eq:DecTerm} would in principle allow for a model-independent analysis of neutrino decoherence effects in oscillation data. However, given the large number of free parameters that it contains, such an analysis is realistically not viable from a practical point of view.
To simplify the theoretical framework, focusing on a $N$-neutrino system, we can expand the operators in Eq~\eqref{eq:DecTerm} in terms of the Gell-Mann matrices from the $SU(N)$ group~\cite{Benatti:2000ph,Gago:2002na,Carrasco:2018sca,Buoninfante:2020iyr,Stuttard:2020qfv}

\begin{equation}
\label{eq:GellMannexp}
\mathcal{D}[\rho_\nu(t)] = c_k \lambda^k ~\left(\mathrm{with}~\rho_\nu=\sum \rho_\nu^k \lambda^k \mathrm{and}~\mathcal{O}_j=\sum \mathcal{O}^j_k \lambda^k \right) \,.
\end{equation} 
In the previous expression, $c_k$ are the coefficients of the expansion, and the index $k$ runs from 0 to $N^2-1$. In the specific case of three neutrino flavors: $k = 0\ldots 8$, $\lambda^0$ is the identity matrix and $\lambda^k$ are the Gell-Mann matrices satisfying $[\lambda^a,\lambda^b]=i\sum_c f_{abc}\lambda^c$, $f_{abc}$ being the structure constants of $SU(3)$.
We can then express the decoherence term as $\mathcal{D}[\rho_\nu(t)] = (\mathbf{D}_{k \ell}~ \rho_\nu^\ell) \lambda^k$, where $\rho_\nu^\ell$ are the coefficients of the neutrino density matrix, expanded in the $SU(N)$ basis (see  Eq.~\eqref{eq:GellMannexp}), and $\mathbf{D}_{k \ell} ~\rho_\nu^\ell = c_k$ are the elements of a $N^2 \times N^2$ matrix representing the free parameters of the system.
Focusing again on a three-neutrino scenario, the matrix $\bf{D}$ can be parameterized by 45 parameters as follows

\begin{equation}
\label{eq:Dmatrixfull}
\bf{D}=
\begin{pmatrix}
-\Gamma_{0} & \beta_{01} & \beta_{02} & \beta_{03} & \beta_{04} & \beta_{05} & \beta_{06} & \beta_{07} & \beta_{08} \\ 
\beta_{01} & -\Gamma_{1} & \beta_{12} & \beta_{13} & \beta_{14} & \beta_{15} & \beta_{16} & \beta_{17} & \beta_{18} \\ 
\beta_{02} & \beta_{12} & -\Gamma_{2} & \beta_{23} & \beta_{24} & \beta_{25} & \beta_{26} & \beta_{27} & \beta_{28} \\ 
\beta_{03} & \beta_{13} & \beta_{23} & -\Gamma_{3} & \beta_{34} & \beta_{35} & \beta_{36} & \beta_{37} & \beta_{38} \\ 
\beta_{04} & \beta_{14} & \beta_{24} & \beta_{34} & -\Gamma_{4} & \beta_{45} & \beta_{46} & \beta_{47} & \beta_{48} \\ 
\beta_{05} & \beta_{15} & \beta_{25} & \beta_{35} & \beta_{45} & -\Gamma_{5} & \beta_{56} & \beta_{57} & \beta_{58} \\ 
\beta_{06} & \beta_{16} & \beta_{26} & \beta_{36} & \beta_{46} & \beta_{56} & -\Gamma_{6} & \beta_{67} & \beta_{68} \\ 
\beta_{07} & \beta_{17} & \beta_{27} & \beta_{37} & \beta_{47} & \beta_{57} & \beta_{67} & -\Gamma_{7} & \beta_{78} \\ 
\beta_{08} & \beta_{18} & \beta_{28} & \beta_{38} & \beta_{48} & \beta_{58} & \beta_{68} & \beta_{78} & -\Gamma_{8} \\
\end{pmatrix}\,,
\end{equation}	
where all entries are real scalars and the $\Gamma_i$  are positive in order to satisfy the relation Tr$(\rho_\nu(t))= 1$.
As already mentioned, while a general model-independent analysis might be contemplated, it proves convenient to reduce the number of free parameters by imposing general physical conditions on $\bf{D}$~\cite{Lisi:2000zt,Benatti:2000ph,Gago:2002na,Oliveira:2010zzd,BalieiroGomes:2018gtd,Stuttard:2020qfv}. These include:
\begin{itemize}
    \item \textit{Unitarity of the system}.  Probability conservation implies ${\bf{D}}_{k0} =  {\bf{D}}_{0\ell}= 0$~\cite{Buoninfante:2020iyr}, given that $f_{ab0} = 0$.
    \item \textit{Complete positivity} of the time-evolution $\rho_\nu$ place conditions on the diagonal elements, thus making them not completely independent.
    \item  \textit{Entropy increase.} The condition $\mathcal{O}_j = \mathcal{O}_j^\dagger$ implies that the Von Neumann entropy $S = -\mathrm{Tr}(\rho_\nu \mathrm{ln}\rho_\nu)$ increases with time~\cite{Benatti:1987dz}.
    \item \textit{Energy conservation} of the neutrino subsystem is satisfied through the commutation relation $[H,\mathcal{O}_j] = 0$. This bound includes the decoherence effect in the evolution.
\end{itemize}

Although in principle energy nonconservation in the neutrino subsystem could be envisaged~\cite{Benatti:2000ph,Oliveira:2010zzd}, while of course always being satisfied for the global system, in the following we work under the requirement of energy conservation on the neutrino subsystem. These assumptions allow for a remarkable simplification of the dissipator in Eq.~\eqref{eq:Dmatrixfull}, which eventually assumes the diagonal form~\cite{BalieiroGomes:2016ykp,Oliveira:2016asf,BalieiroGomes:2018gtd}

\begin{equation}
 \label{eq:Dmatrixdiag}
	{\bf{D}}= -\mathrm{diag}(\Gamma_{21},\Gamma_{21},0,\Gamma_{31}, \Gamma_{31},\Gamma_{32},\Gamma_{32},0) \,,
\end{equation}
in the case of three neutrino flavors. Comparing with Eq.~\eqref{eq:Dmatrixfull}, we see that imposing the above-mentioned conditions in practice leads to $\beta_{ij}=0$ and $\Gamma_1 = \Gamma_2 \equiv \Gamma_{21}$, $\Gamma_4 = \Gamma_5 \equiv \Gamma_{31}$ and $\Gamma_6 = \Gamma_7 \equiv \Gamma_{32}$. In principle, also the so-called relaxation parameters $\Gamma_{3}$ and $\Gamma_{8}$ could be present in Eq.~\eqref{eq:Dmatrixdiag}. However, since the terrestrial experiments that we consider here are not expected to have competitive sensitivity with bounds from solar neutrinos~\cite{deHolanda:2019tuf}, we set $\Gamma_3 = \Gamma_8 = 0$ throughout the paper.

As anticipated, the interaction of the neutrino subsystem with the environment, through the matrix Eq.~\eqref{eq:Dmatrixdiag} will induce damping effects of the form $e^{-\Gamma_{ij} L}$ in the oscillation probabilities, $L$ being the experiment baseline and $\Gamma_{ij}$ quantifying the strength of the damping effects, resulting in a coherence length $L_\mathrm{coh} = 1/\Gamma_{ij}$. While remaining agnostic of the (quantum-gravity) origin of such dissipating terms, we follow a phenomenological approach and introduce a general form for the energy-dependence of the dissipator matrix controlling the decoherence effects. We assume a power-law energy-
dependence following previous studies in the literature~\cite{Lisi:2000zt,Fogli:2007tx,Coloma:2018idr,Guzzo:2014jbp,Carrasco:2018sca,Gomes:2020muc,Stuttard:2020qfv}

\begin{equation}
\label{eq:gamma_E}
\Gamma_{ij}(E) = \Gamma_{ij}(E_0) \left( \frac{E}{E_0} \right)^n \, ,
\end{equation}

where $E_0$ is a pivot energy scale which we set to $E_0 = 1$~GeV, as previously done in the literature, and $n$ is a power-law index to be tested experimentally. Given the lack of a fundamental theory for quantum gravity, the dependence of $\Gamma_{ij}$ on the neutrino energy is currently unknown. We will make the hypothesis that it can assume the following integer values: $n = [-2, -1, 0, +1, +2]$. Note that this power-law dependence on the neutrino energy breaks Lorentz invariance, except for the case with $n = -1$ which imitates the oscillation energy dependence. 
Our choice for $n$ thus accommodates: a case where decoherence might be
induced by lightcone fluctuations~\cite{Stuttard:2021uyw} ($n=-2$); a Lorentz-invariant
case ($n= -1$)~\cite{Lisi:2000zt};
the energy-independent case ($n = 0$); $n = +1$ which mimics a cross section-like energy-dependence, or as suggested in Ref.~\cite{Liu:1997km}; and the case ($n=+2$) that can arise in quantum-gravity models, in which $\Gamma_i(E_0)\sim
O(M_{\rm Planck}^{-1})\sim 10^{-19}$~GeV is expected~\cite{Ellis:1995xd,Ellis:2000dy,Gambini:2003pv,Ellis:1996bz, Ellis:1997jw}. 
For the sake of completeness, let us also recall that the loss of coherence due to wave-packet separation would correspond to the case $n=-4$. While we do not cover this scenario here, we refer the reader to Refs.~\cite{deGouvea:2020hfl,deGouvea:2021uvg} for recent analyses. We also note that the $n=-1$ case is phenomenologically similar to invisible neutrino decay~\cite{Stuttard:2020qfv}. Finally, it has recently been suggested that scenarios with an extreme energy dependence ($n\leq-10$) could explain the Gallium anomaly~\cite{Farzan:2023fqa} without the need of light sterile neutrinos and hence without entering in conflict with bounds from other short-baseline experiments~\cite{Giunti:2022btk}.

\subsection{Phenomenological models}
\label{sec:models}

\begin{table}[!h]
\centering
\begin{tabular}{ c | c} 
\hline
Model A ~~~&~~~ We vary $\Gamma_{21}=\Gamma_{31}=\Gamma_{32}$\\
Model B ~~~&~~~ We vary $\Gamma_{21}=\Gamma_{31}$ and keep $\Gamma_{32}= 0$\\
Model C ~~~&~~~ We vary $\Gamma_{21}=\Gamma_{32}$ and keep $\Gamma_{31}= 0$\\
Model D ~~~&~~~ We vary $\Gamma_{31}=\Gamma_{32}$ and keep $\Gamma_{21}= 0$\\
Model E ~~~&~~~ We vary $\Gamma_{21}$ and keep $\Gamma_{31}=\Gamma_{32}= 0$\\
Model F ~~~&~~~ We vary $\Gamma_{31}$ and keep $\Gamma_{21}=\Gamma_{32}= 0$\\
Model G ~~~&~~~ We vary $\Gamma_{32}$ and keep $\Gamma_{21}=\Gamma_{31}= 0$
\\\hline
\end{tabular}
\caption{The phenomenological models considered in this paper.}
\label{tab:models}
\end{table} 

The choice of {\bf{D}} defines the flavor structure of the decoherence effects, and ultimately the flavor ratio of a  neutrino ensemble once coherence is completely lost. In the analyses of this paper we consider several scenarios, reflected in activation of different $\Gamma_{ij}$ simultaneously. The models that we consider are summarized in Table~\ref{tab:models}, and can be grouped into three categories; all three $\Gamma_{ij}$ activated (A), two $\Gamma_{ij}$ activated (B, C, D), and only a single $\Gamma_{ij}$ activated (E, F, G). In each case we test each of the above mentioned energy dependencies.

Activating differing $\Gamma_{ij}$ results in differing matrix elements of the $[H,\rho_\nu(t)]$ term in Eq.~\eqref{eq:rho_time_dep} being damped. In particular, specific active pairs of $\Gamma_{ij}$ damp differing oscillations frequencies resulting from the solar ($\Delta m^2_{21}$) and atmospheric ($\Delta m^2_{3j}$) mass splittings. Model D for instance results in the (faster) atmospheric sector oscillations being damped whilst the (slower) solar sector oscillations are not. Such a scenario could therefore represent new physics that depends on the mass splittings (or equivalently oscillation wavelengths), such as lightcone fluctuations~\cite{Stuttard:2021uyw} or a mass-dependent coupling. Model A represents the case where all oscillation channels are damped together, and can result from perturbations to the neutrino wavefunction phase~\cite{Stuttard:2020qfv}. In general, combining experimental data from various oscillation channels/experiments will be required to optimally test all models.

\subsection{Neutrino oscillation probability and decoherence}
\label{sec:osc_prob}

The neutrino oscillation probability in presence of the dissipator under consideration here, Eq.~\eqref{eq:Dmatrixdiag}, has been derived many times in the literature, see for example Ref.~\cite{BalieiroGomes:2018gtd}. The result reads

\begin{align}
    P(\nu_\alpha\to\nu_\beta)=\delta_{\alpha\beta} &- 2\sum_{k>j}\Re\left[ \tilde{U}^*_{\alpha k}\tilde{U}_{\beta k}\tilde{U}_{\beta j}\tilde{U}^*_{\beta j} \right] ~\left[1-\cos \left(\frac{\Delta \tilde{m}^2_{kj}L}{2E}\right)~e^{-\Gamma_{kj}(E)L}\right]\nonumber
    \\
    &+2\sum_{k>j}\Im\left[ \tilde{U}^*_{\alpha k}\tilde{U}_{\beta k}\tilde{U}_{\beta j}\tilde{U}^*_{\beta j} \right] ~\sin \left(\frac{\Delta \tilde{m}^2_{kj}L}{2E}\right)~e^{-\Gamma_{kj}(E)L}\,,
    \label{eq:osc_prob}
\end{align}
where $\Gamma_{ij}(E)$ is given in Eq.~\eqref{eq:gamma_E} and $\tilde{U}$ and $\Delta\tilde{m}^2_{kj}$ are the mixing matrix and mass splittings in matter. In our analyses we calculate these quantities using the expressions provided in Ref.~\cite{Denton:2016wmg}, which currently constitute the most precise approximation for neutrino oscillations in constant matter~\cite{Barenboim:2019pfp}. In the following, matter effects will be relevant in the analyses of KamLAND, NOvA and DUNE. Note that the expression of the neutrino oscillation probability is equivalent to the standard expression when $\Gamma_{ij}(E)\to 0$ (i.e. when no decoherence effects are present). It should be noted that Eq.~\eqref{eq:osc_prob} is an approximation, since when including matter effects non-diagonal elements can appear in ${\bf{D}}$ after rotation to the matter basis~\cite{Carpio:2017nui}. However, this approximation is valid for the analyses performed in this paper, because of several reasons: Firstly, matter effects are never very strong as they would be, for example, in an analysis of atmospheric neutrinos. Secondly, most of the experiments considered here are mainly using disappearance channels, which are much less affected by matter effects than appearance channels. In experiments that use both disappearance and appearance channels, the statistics of the appearance channels are much smaller than those of the disappearance channels. Additionally, the non-diagonal terms would only effect the neutrino-oscillation probability at large energies in the tails of the spectra which do not have many events.
\begin{figure}
\begin{subfigure}{0.48\textwidth}
    \includegraphics[width=\textwidth]{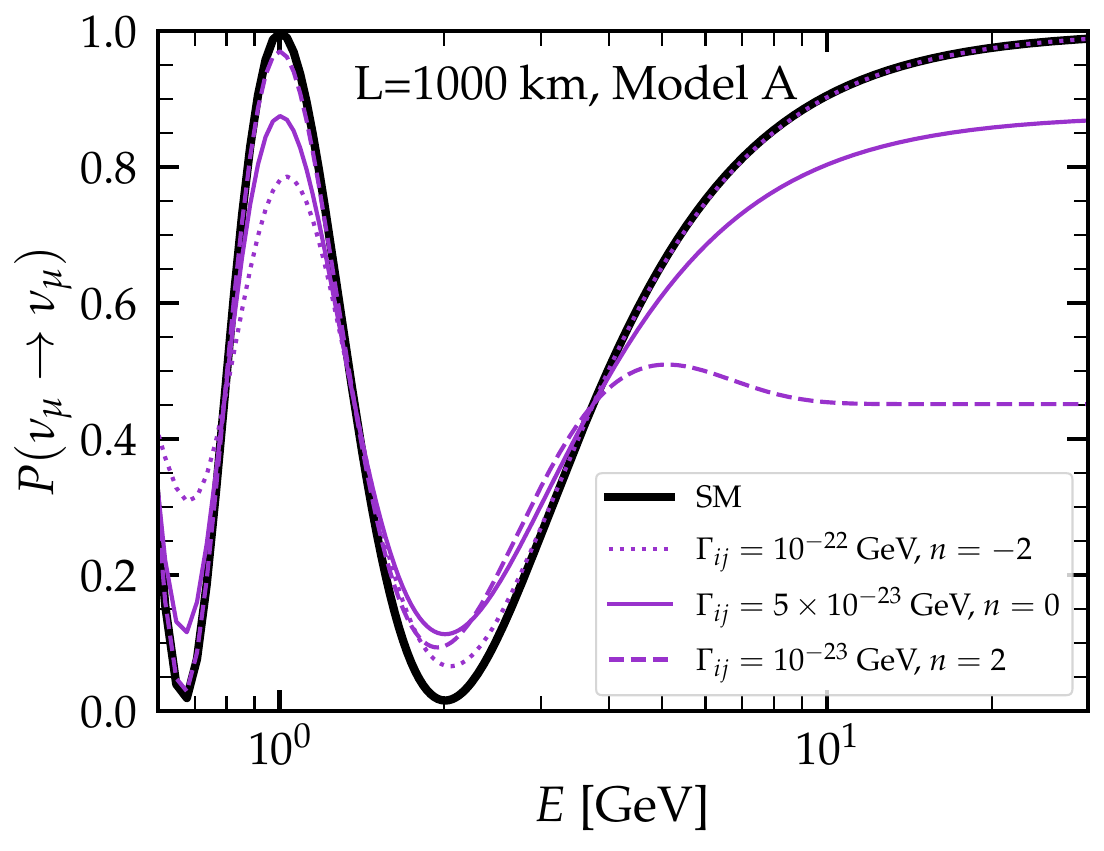}
\end{subfigure}
\hfill
\begin{subfigure}{0.49\textwidth}
    \includegraphics[width=\textwidth]{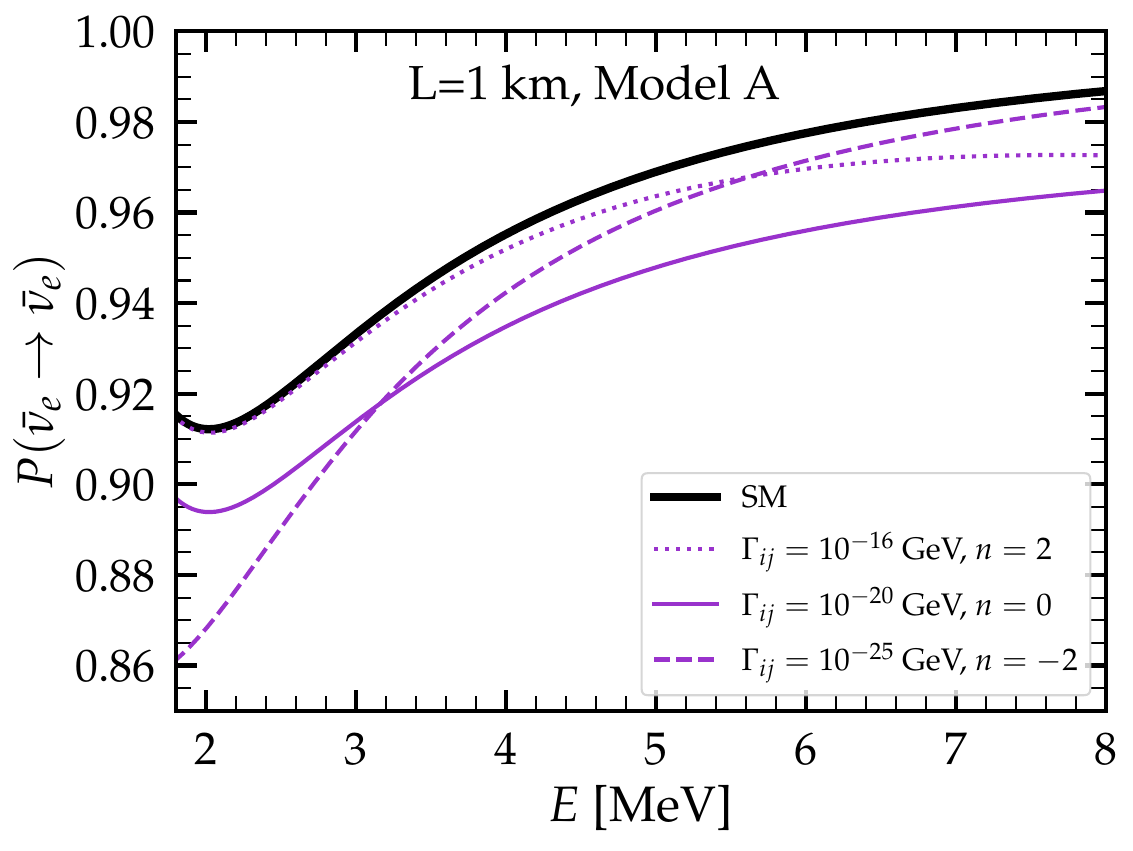}
\end{subfigure}
\hfill
\caption{Neutrino oscillation probabilities for the baselines, energies and flavors relevant for several of the experiments considered in this paper, in the case of Model A and for various $n$. The left panel is characteristic of long-baseline accelerator neutrino experiments, whilst the right panel represents reactor neutrino experiments.}
\label{fig:osc1}
\end{figure}

\begin{figure}
\begin{subfigure}{0.48\textwidth}
    \includegraphics[width=\textwidth]{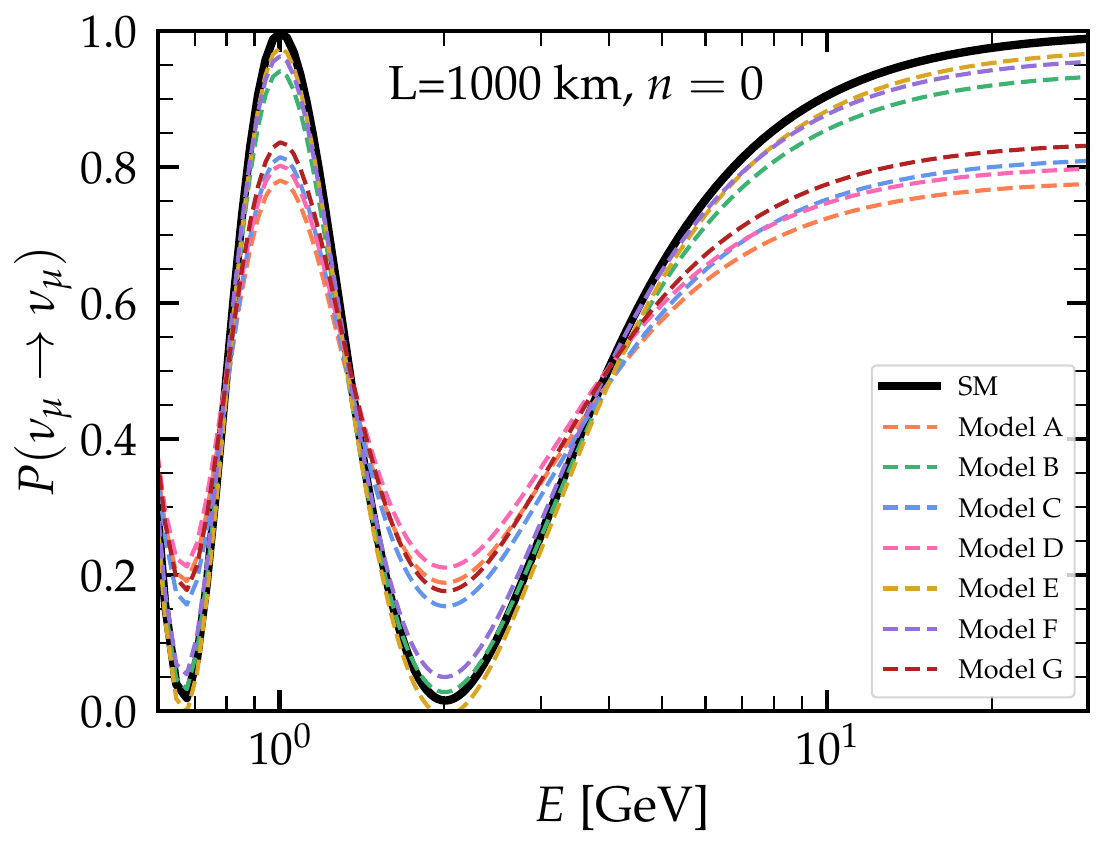}
\end{subfigure}
\hfill
\begin{subfigure}{0.49\textwidth}
    \includegraphics[width=\textwidth]{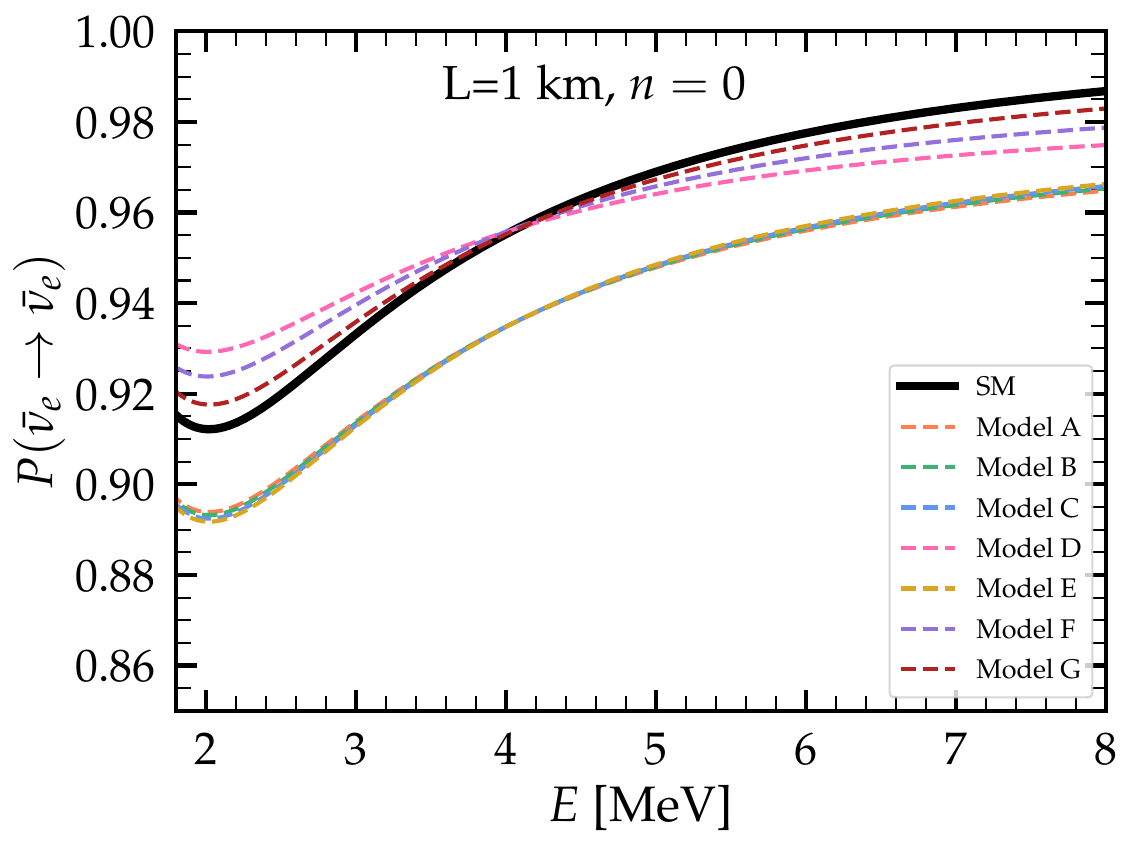}
\end{subfigure}
\caption{Neutrino oscillation probabilities for the baselines, energies and flavors relevant for several of the experiments considered in this paper, for all phenomenological models (see Table~\ref{tab:models}). The left panel is characteristic of long-baseline accelerator neutrino experiments, whilst the right panel represents reactor neutrino experiments. We fix $n=0$ and $\Gamma_{ij}=10^{-22}$ GeV (left panel), $\Gamma_{ij}=10^{-20}$ GeV (right panel, models A-B-C-E) and $\Gamma_{ij}=10^{-19}$ GeV (right panel, models D-F-G).}
\label{fig:osc2}
\end{figure}
We show in Fig.~\ref{fig:osc1} the $\nu_\mu$ (left panel) and $\bar{\nu}_e$ (right panel) survival probabilities using baselines and energies relevant for the experiments considered in this paper. As can be seen, different choices of $n$ can induce different forms of spectral distortions. We fix $\Gamma_{ij}(E)$ to values in the ballpark of current bounds, which we will infer in Sec.~\ref{sec:results}. Experiments using low-energy neutrinos will be more sensitive to scenarios with negative $n$, while high-energy neutrinos will allow us to set stronger bounds when $n>0$, as it is particularly clear from the left plot.
In Fig.~\ref{fig:osc2} we keep $n=0$ fixed and show the impact of the different models on the oscillation probabilities. From the right panel (representing reactor experiments) we see that the models can be separated into two groups: One group, which has $\Gamma_{21}$ activated and one which has not. If $\Gamma_{21}$ is active we obtain, in addition to possible spectral distortions, also an overall shift of the oscillation probability. Note that a simple shift might not result in an observable effect if one only takes ratios of events at different detectors. Therefore we expect the reactor experiments to have best sensitivity to models which have $\Gamma_{21}$ and at least one other $\Gamma$ activated, while also using $n<0$.
Also in the case of long-baseline accelerator experiments (left panel) we find that there exist two groups of models. One group which has $\Gamma_{32}$ active, and one without. As can be seen from the left panel of Fig.~\ref{fig:osc2}, the deviation from the standard three-neutrino oscillation probability is particularly large for the former models (A-C-D-G). We can therefore expect accelerator experiments to have best sensitivity to models with (at least) $\Gamma_{32} \neq 0$ and positive $n$. Of course, this expectation based on oscillation plots is somewhat simplified, since in the real analyses correlations with standard neutrino oscillation parameters and systematic uncertainties of the experiments must be taken into account. Nevertheless, as will be shown later, the strengths of the bounds for the models under consideration will follow this simplified expectation in many cases.

\section{Experiments included in the analysis}
\label{sec:experiments}

In this section we discuss the experimental data that have been considered in this paper and give some details of the analyses performed. We have used data from several reactor and long-baseline accelerator experiments, discussed in Sections~\ref{sec:reacs} and~\ref{sec:accels}, respectively. We also detail in Section~\ref{sec:future_exps} the analysis procedure for the expected sensitivities at the next-generation experiments JUNO and DUNE. It should be noted that the decoherence effects discussed in this paper do not affect solar neutrino measurements. We will therefore always impose a prior on the solar mixing angle, fixing $\sin^2\theta_{12} = 0.318\pm0.016$~\cite{deSalas:2020pgw}.

\subsection{Reactor experiments: KamLAND, Daya Bay and RENO}
\label{sec:reacs}

KamLAND was a reactor experiment which was placed at the site of the former Kamiokande experiment and was used to measure electron antineutrinos from more than 50 reactors, located at distances ranging from $\sim100$~km to $\sim1000$~km. 
Due to the long-baselines, matter effects are not negligible at KamLAND, and must be included in the calculation of the neutrino oscillation probability.
For the present analysis we include the data presented in Refs.~\cite{Gando:2010aa,kamland_web}.
The $\chi^2$ function used for KamLAND reads

\begin{equation}
 \chi^2_\text{KL}(\vec{p}) = \min_{\vec{\alpha}}\left\{\sum_{i=1}^{N_\text{KL}}\left(\frac{N_{\text{dat},i} - N_{\text{exp},i}(\vec{p},\vec{\alpha})}{\sigma^\text{KL}_i}\right)^2 + \frac{(N^{\text{tot}}_{\text{dat}} - N^{\text{tot}}_{\text{exp}}(\vec{p},\vec{\alpha}))^2}{N^{\text{tot}}_{\text{dat}}}
+ 
 \sum_k \left(\frac{\alpha_k - \mu_k}{\sigma_k} \right)^2\right\}+\chi^2_{\theta_{12}}\,,
\end{equation}
where the index $i$ runs over the energy bins.
In the previous expression, $N_{\text{dat},i}$ denote the observed event numbers per energy bin, while $N_{\text{exp},i}(\vec{p},\vec{\alpha})$ indicate the expected event numbers for a given set of oscillation parameters $\vec{p}$ and systematic uncertainties $\vec{\alpha}$.
The second term is a penalty term on the total number of events, 
while the third term contains penalty factors for all of the systematic uncertainties $\alpha_k$, with expectation value $\mu_k$ and standard deviation $\sigma_k$. 
The systematic uncertainties that are included in our analysis account for reactor uncertainties (normalizations related to the different reactors and an uncorrelated shape error) and detector uncertainties (detection efficiency and energy scale). The last term contains the above mentioned penalty for $\sin^2\theta_{12}$.
We use GLoBES~\cite{Huber:2004ka,Huber:2007ji} to compute the event numbers and to perform the statistical analysis.
The reactor fluxes are parameterized as in Ref.~\cite{Huber:2004xh} and the inverse beta-decay cross section is taken from Ref.~\cite{Vogel:1999zy}. 
We have found that the measurement of $\Delta m_{21}^2$ obtained from the analysis of KamLAND data is robust under the decoherence effects discussed here. We will therefore include another penalty when marginalizing over $\Delta m_{21}^2$ in all experiments discussed in the following subsections, corresponding to $\Delta m_{21}^2 = (7.53\pm0.22)\times10^{-5}$~eV$^2$.\\

Apart from KamLAND, we also include data from the reactor experiments Daya Bay and RENO. We analyze the data sets corresponding to 2900 days of running time at RENO~\cite{jonghee_yoo_2020_4123573} and 1958 days of running time at Daya Bay~\cite{DayaBay:2018yms}. 
In the statistical analyses, we include uncertainties related to the thermal power for each core, to the detection efficiencies, uncertainties on the fission fractions, a shape uncertainty for each energy bin, and an uncertainty on the energy scale.
The $\chi^2$ function for RENO is
\begin{equation}
\label{eq:RENOchi2}
 \chi^2_\text{RENO}(\vec{p}) = \min_{\vec{\alpha}}\left\{\sum_{i=1}^{N_\text{RENO}}\left(\frac{R^{F/N}_{\text{dat},i} - R^{F/N}_{\text{exp},i}(\vec{p},\vec{\alpha})}{\sigma^\text{RENO}_i}\right)^2 + \sum_k \left(\frac{\alpha_k - \mu_k}{\sigma_k} \right)^2\right\} + \chi^2_{\text{solar}}\,.
\end{equation}
Here, $R^{F/N}_i = F_i/N_i$, is the ratio of events at the far ($F_i$) and near ($N_i$) detectors in the $i$th energy bin. 
In particular, $R_{\text{dat},i}$ are the background-subtracted observed event ratios, while $R_{\text{exp},i}(\vec{p},\vec{\alpha})$ are the expected event ratios for a given set of oscillation parameters $\vec{p}$. 
The uncertainty for each bin is given by $\sigma^\text{RENO}_i$.
The second term contains penalty factors for all the systematic uncertainties $\alpha_k$, again with expectation value $\mu_k$ and standard deviation $\sigma_k$. The last term $\chi^2_{\text{solar}}$ contains the solar penalty for $\sin^2\theta_{12}$ and the KamLAND penalty for $\Delta m_{21}^2$.  Similarly, for Daya Bay, we define

\begin{align}
\label{eq:DBchi2}
 \chi^2_\text{DB}(\vec{p}) = &\min_{\vec{\alpha}}\left\{\sum_{i=1}^{N_\text{DB}}\left(\frac{R^{F/N_1}_{\text{dat},i} - R^{F/N_1}_{\text{exp},i}(\vec{p},\vec{\alpha})}{ \sigma_i^{F/N_1}}\right)^2  + \sum_{i=1}^{N_\text{DB}}\left(\frac{R^{N_2/N_1}_{\text{dat},i} - R^{N_2/N_1}_{\text{exp},i}(\vec{p},\vec{\alpha})}{\sigma_i^{N_2/N_1}}\right)^2 \right.\nonumber
 \\
&\left. +  \sum_k \left(\frac{\alpha_k - \mu_k}{\sigma_k} \right)^2\right\} + \chi^2_{\text{solar}}\,.
\end{align}
Here, we take the ratios between the far and the first near detector and between the two near detectors. 
We use the same reactor flux model and inverse beta-decay cross section as in the analysis of KamLAND. 
%

\subsection{Accelerator experiments: MINOS/MINOS+, T2K and NOvA}
\label{sec:accels}

MINOS was an accelerator-based neutrino oscillation experiment studying muon neutrinos produced at Fermilab at the NuMI beam facility and detected at two detectors located at 1.04~km and 735~km away from the source. First, during the MINOS phase, the neutrino beam peaked at an energy of 3~GeV. Later, during the MINOS+ phase, the energy of the beam was increased, peaking at 7~GeV. Here we consider data corresponding to an exposure of $10.56\times10^{20}$ proton-on-target (POT) in MINOS and $5.80\times10^{20}$ POT in MINOS+, collected in the same detectors~\cite{Adamson:2017uda}. We adopted the analysis procedure followed by the experimental collaboration for the search of active-sterile neutrino oscillations in Ref.~\cite{Adamson:2017uda}, by modifying the public MINOS/MINOS+ code to account for decoherence effects instead of  active-sterile oscillations. The statistical analysis is performed by using the following $\chi^2$ definition: 

\begin{equation}
    \chi^2_{\mathrm{MINOS}}(\vec{p}) = \sum_{\text{Det.}}\sum_{i,j} (N_{\text{dat},i} - N_{\text{exp},i}(\vec{p}))~[V_{\text{cov.}}]^{-1}_{ij}~(N_{\text{dat},j} - N_{\text{exp},j}(\vec{p})) + \chi^2_{\text{solar}}\,,
\end{equation}
where $N_{\text{dat},i}$ and $N_{\text{exp},i}$ are the observed and the predicted event rates, in the energy bin $i$. The covariance matrix $V_\text{cov.}$ accounts for the contributions from several sources of systematic uncertainties~\cite{Adamson:2017uda}, and the first sum runs over the two detectors. The last term, $\chi^2_{\text{solar}}$, is the same as in Eqs.~\eqref{eq:RENOchi2} and \eqref{eq:DBchi2}.

We further consider data collected by T2K~\cite{Abe:2021gky} and NOvA~\cite{NOvA:2021nfi}. The T2K collaboration observed events induced by neutrinos and antineutrinos, corresponding to an exposure at Super-Kamiokande of 1.97$\times10^{21}$ POT in neutrino mode and 1.63$\times10^{21}$ POT in antineutrino mode. 
NOvA has instead reached 13.6$\times10^{20}$~POT in neutrino mode~\cite{NOvA:2018gge} and 12.5$\times10^{20}$~POT in antineutrino mode.
For the energy reconstruction in both experiments we assume Gaussian smearing adding bin-to-bin efficiencies, which are adjusted to reproduce the best-fit spectra reported by the experimental collaborations.
Our statistical analysis includes  several sources of systematic uncertainties, related to the signal and background predictions. 
The $\chi^2$ function for the analysis of T2K and NOvA data (and DUNE, as we will see later) is given by
\begin{equation}
\label{eq:T2KNOvAchi2}
 \chi^2_{\mathrm{T/N/D}}(\vec{p})=\min_{\vec{\alpha}}
 \sum_\text{channels}2\sum_i \left[ N_{\text{exp},i}(\vec{p},\vec{\alpha})- N_{\text{dat},i} +
 N_{\text{dat},i} \log \left(\frac{N_{\text{dat},i}}{N_{\text{exp},i}(\vec{p},\vec{\alpha})}\right)\right] 
 + \sum_i \left(\frac{\alpha_i}{\sigma_i}\right)^2 + \chi^2_{\text{solar}}\,,
\end{equation}
where again $N_{\text{dat},i}$ and $N_{\text{exp},i}$ are the data and prediction in bin $i$ and the first sum is taken over the different oscillation channels: $\nu_{\mu}\to\nu_{\mu}$, $\overline{\nu}_{\mu}\to\overline{\nu}_{\mu}$, $\nu_{\mu}\to\nu_e$ and $\overline{\nu}_{\mu}\to\overline{\nu}_e$. The last two terms are equivalent to all other cases.

\subsection{Future experiments: JUNO and DUNE}
\label{sec:future_exps}

In this subsection we discuss the details for the sensitivity analyses at JUNO and DUNE. When performing these sensitivity analyses we generate a mock data set for both experiments using the neutrino oscillation parameters from Ref.~\cite{deSalas:2020pgw} and without decoherence effects. In the statistical analysis we vary the decoherence parameters and marginalize over the standard parameters, equivalently to the other experiments.

JUNO~\cite{JUNO:2015zny} is a next-generation reactor experiment, which aims to precisely measure the solar neutrino oscillation parameters, the atmospheric mass splitting, and the neutrino mass ordering.
JUNO will consist of eight reactors~\cite{IceCube-Gen2:2019fet,JUNO:2021vlw} located at around 52~km distance from the main detector.
Our simulation of the experiment follows the descriptions in Refs.~\cite{JUNO:2015zny,IceCube-Gen2:2019fet,JUNO:2021vlw,Forero:2021lax} assuming 8 years of running time.
Our statistical analyses are performed with

\begin{equation}
 \chi^2_{\mathrm{JUNO}}(\vec{p}) = \min_{\vec{\alpha}}\left\{\sum_{i=1}^{N_\text{JUNO}}\left(\frac{N_{\text{dat},i} - N_{\text{exp},i}(\vec{p},\vec{\alpha})}{\sigma_i}\right)^2 + \sum_k \left(\frac{\alpha_k - \mu_k}{\sigma_k} \right)^2\right\}  + \chi^2_{\text{solar}}\,,
\end{equation}
where the definition of the quantities is equivalent to the other cases, barring that this time $N_{\text{dat},i}$ consists of the mock data set.\\

DUNE~\cite{DUNE:2020lwj,DUNE:2020ypp,DUNE:2020mra,DUNE:2020txw}~ is going to be the successor experiment of NOvA. It will also consist of two detectors, which are going to be exposed to a megawatt-scale neutrino beam produced at
Fermilab, composed of (nearly) only muon neutrinos or antineutrinos. The
near detector will be placed approximately 600 meters away from the
source of the beam, while the far detector, divided into four
modules, each using 10~kton of argon as detection material, will be
installed 1285 kilometres away, deep underground at the Sanford
Underground Research Facility in South Dakota.
To simulate the neutrino signal at DUNE we use the latest configuration file for GLoBES provided by the DUNE collaboration \cite{DUNE:2021cuw}, which assumes 6.5 years of running time in both neutrino and antineutrino modes. 
Our analysis includes disappearance and appearance channels, simulating both signals and backgrounds. The simulated backgrounds include contamination of antineutrinos (neutrinos) in the neutrino (antineutrino) mode, and also misinterpretation of flavors. 
In addition, we also consider the DUNE-High Energy (HE) flux configuration, which has been proposed to optimize the study of tau neutrinos at DUNE~\cite{DUNEHE}. In this option the beam will peak at higher energies in comparison to the nominal configuration.
When considering the sensitivity of DUNE we will compare the sensitivity using the standard DUNE beam with the high energy beam~\cite{DUNEHE}.
The $\chi^2$ function for DUNE is the same as for T2K and NOvA, see Eq.~\eqref{eq:T2KNOvAchi2}. 

We conclude by summarizing in Table~\ref{tab:exps} some of the experimental details of the experiments considered in this paper.

\noindent
\begin{table}[!h]
\centering
\begin{threeparttable}
\resizebox{\linewidth}{!}{
\begin{tabular}{cccc} 
\toprule
\textbf{Experiment} & \textbf{Baseline} & \textbf{Energy range} & \textbf{Main oscillation channel}\\ 
\midrule
\textcolor{cadmiumgreen}{KamLAND}~\cite{Gando:2010aa} & $\mathcal{O}(100)-\mathcal{O}(1000)$~km & $1.8-8.0$~MeV &~~~  $\overline{\nu}_e\to\overline{\nu}_e$ \\
\textcolor{denim}{Daya Bay}~\cite{DayaBay:2018yms} and \textcolor{orange(colorwheel)}{RENO}~\cite{jonghee_yoo_2020_4123573} & $\mathcal{O}(100)-\mathcal{O}(1000)$~m & $1.8-8.0$~MeV &~~~  $\overline{\nu}_e\to\overline{\nu}_e$ 
\\
\textcolor{purple(x11)}{T2K}~\cite{Abe:2021gky} & 295~km & $0.2-2.0$~GeV &~~~  $\nu_\mu(\overline{\nu}_\mu)\to\nu_\mu(\overline{\nu}_\mu)$  and $\nu_\mu(\overline{\nu}_\mu)\to\nu_e(\overline{\nu}_e)$
\\
\textcolor{darkturquoise}{NOvA}~\cite{NOvA:2021nfi} & 812~km & $0.8-5.0$~GeV &~~~  $\nu_\mu(\overline{\nu}_\mu)\to\nu_\mu(\overline{\nu}_\mu)$  and $\nu_\mu(\overline{\nu}_\mu)\to\nu_e(\overline{\nu}_e)$
\\
\textcolor{red(ncs)}{MINOS/MINOS+}~\cite{Adamson:2017uda} & 735~km & $0-40.0$~GeV &~~~  $\nu_\mu(\overline{\nu}_\mu)\to\nu_\mu(\overline{\nu}_\mu)$  \\
\midrule
\textcolor{ruddypink}{JUNO}~\cite{JUNO:2021vlw} & $\sim53$~km & $1.8-8.0$~MeV & ~~~ $\overline{\nu}_e\to\overline{\nu}_e$ 
\\
\textcolor{slateblue}{DUNE}~\cite{DUNE:2021cuw} & 1285~km & $0.5-20$~GeV &~~~  $\nu_\mu(\overline{\nu}_\mu)\to\nu_\mu(\overline{\nu}_\mu)$  and $\nu_\mu(\overline{\nu}_\mu)\to\nu_e(\overline{\nu}_e)$
\\
\textcolor{airforceblue}{DUNE HE}~\cite{DUNEHE} & 1285~km & $0.5-20$~GeV &~~~ $\nu_\mu(\overline{\nu}_\mu)\to\nu_\mu(\overline{\nu}_\mu)$  and $\nu_\mu(\overline{\nu}_\mu)\to\nu_e(\overline{\nu}_e)$
\\
\hline
\bottomrule
\end{tabular}}
\caption{Details of the experiments considered in this paper. Note that even though the range for DUNE and DUNE-HE is the same, the high-energy beam is much broader and peaks at larger energies resulting in different sensitivities.}
\label{tab:exps}
\end{threeparttable}
\end{table}

\section{Results}
\label{sec:results}
In this section we discuss the results from the analyses performed. Given the form of the neutrino oscillation probability provided in Section~\ref{sec:osc_prob} we expect  a long-baseline experiment to be especially suited to bound the decoherence effects discussed in this paper. Note, however, that if the baseline is too long, oscillations can become fully averaged and any sensitivity to decoherence is lost. This explains why solar neutrino experiments have no sensitivity to the decoherence parameters of interest in our analyses\footnote{On the other hand, solar neutrinos can still be used to bound the relaxation parameters~\cite{deHolanda:2019tuf} although we do not consider them here.}. Due to our choice of $E_0$ in Eq.~\eqref{eq:gamma_E} experiments with low (high) energy neutrinos will be sensitive to models with negative (positive) energy dependence (encoded in the parameter $n$). 

\begin{figure}
\centering
\begin{subfigure}{0.49\textwidth}
    \includegraphics[width=\textwidth]{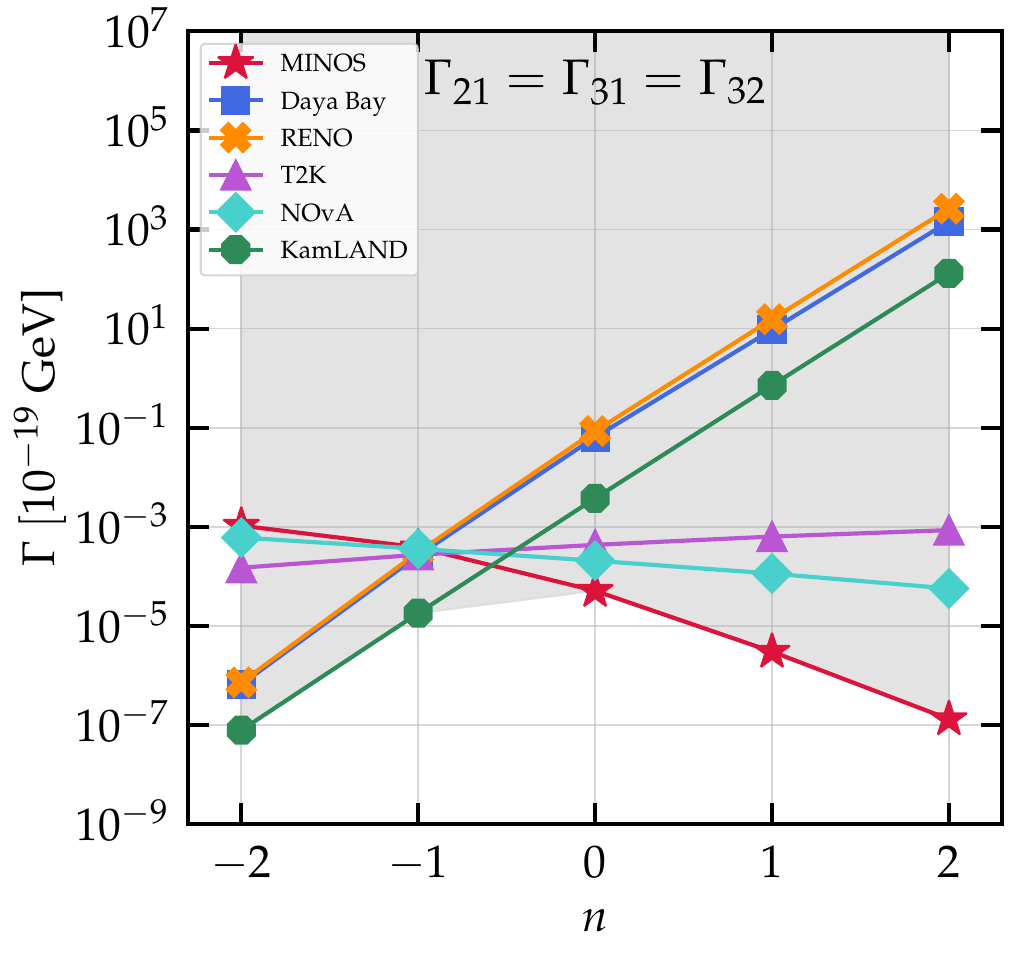}
\end{subfigure}
\hfill
\begin{subfigure}{0.49\textwidth}
    \includegraphics[width=\textwidth]{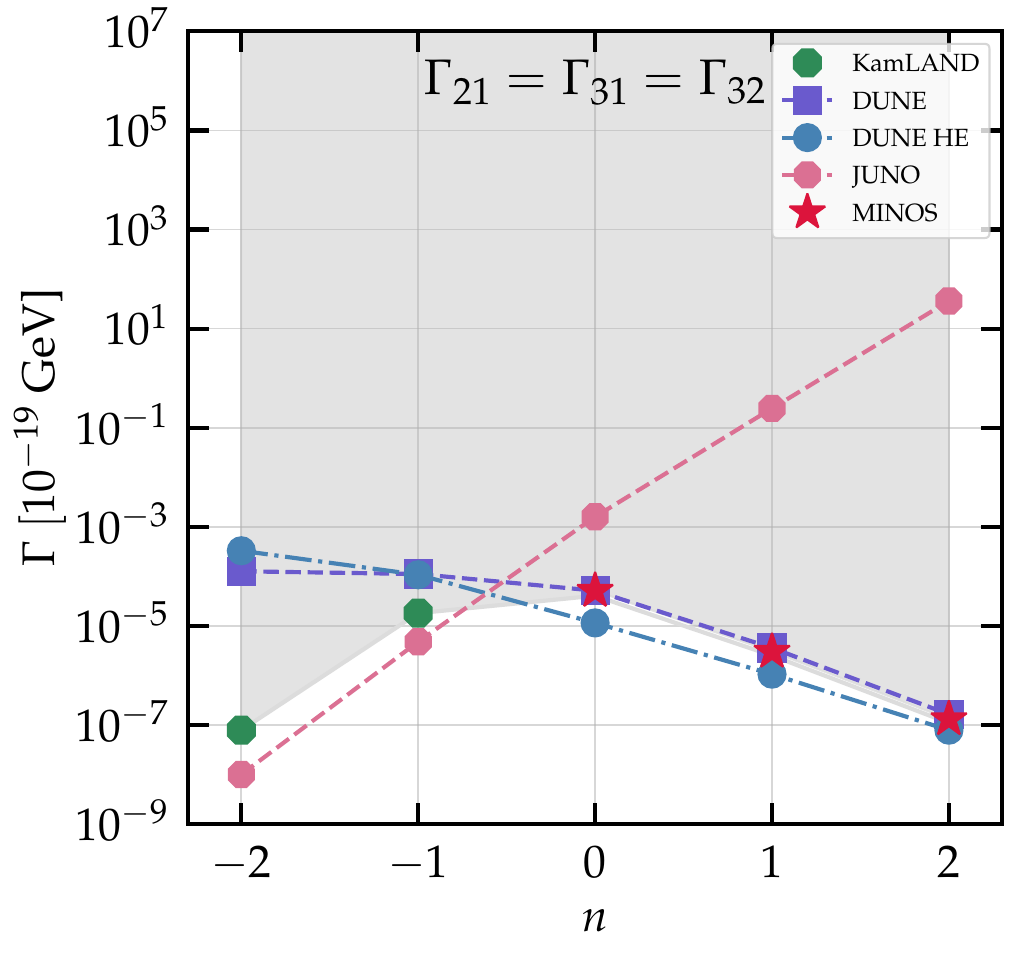}
\end{subfigure}
\caption{Current 90\% C.L. upper limits  (left) and future sensitivities for Model A ($\Gamma \equiv \Gamma_{21}=\Gamma_{31}=\Gamma_{32}$), as a function of $n$ (see Eq.~\eqref{eq:gamma_E}).}
\label{fig:all_gam}
\end{figure}

\begin{figure}
\centering
\begin{subfigure}{0.49\textwidth}
    \includegraphics[width=\textwidth]{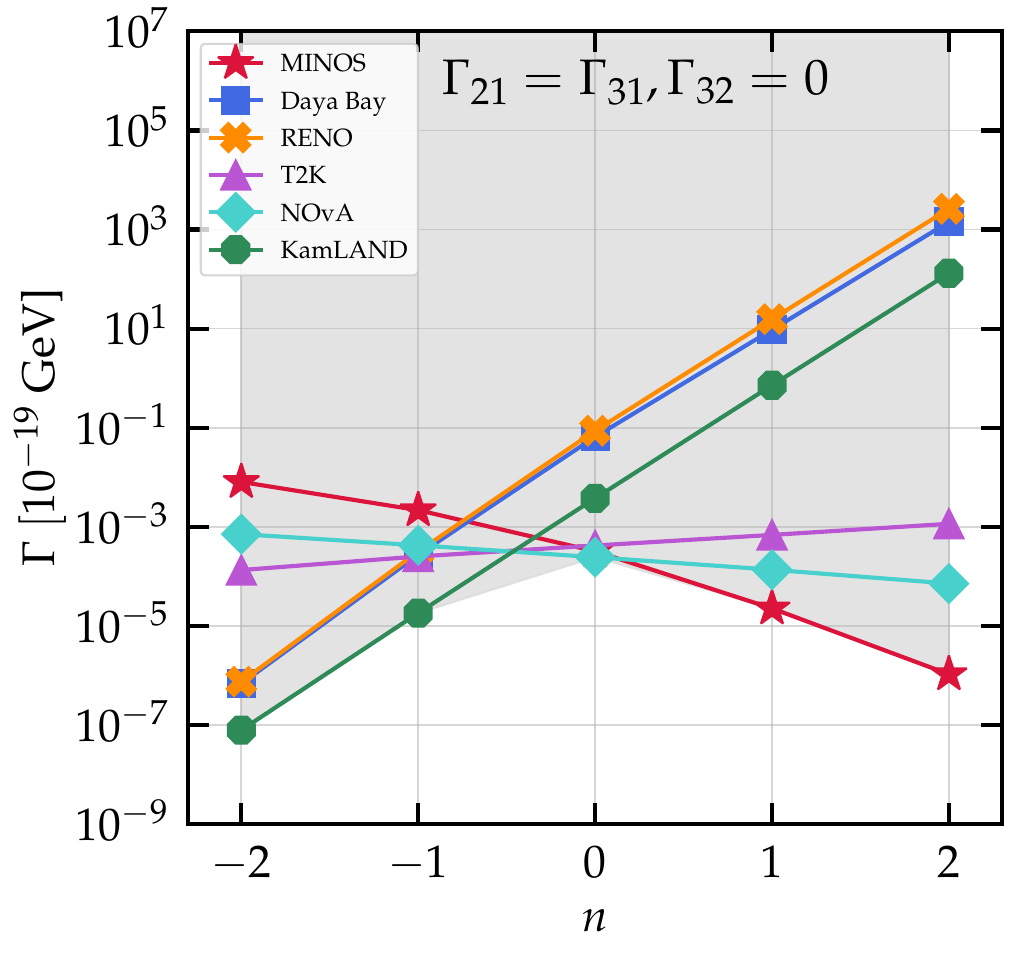}
\end{subfigure}
\hfill
\begin{subfigure}{0.49\textwidth}
    \includegraphics[width=\textwidth]{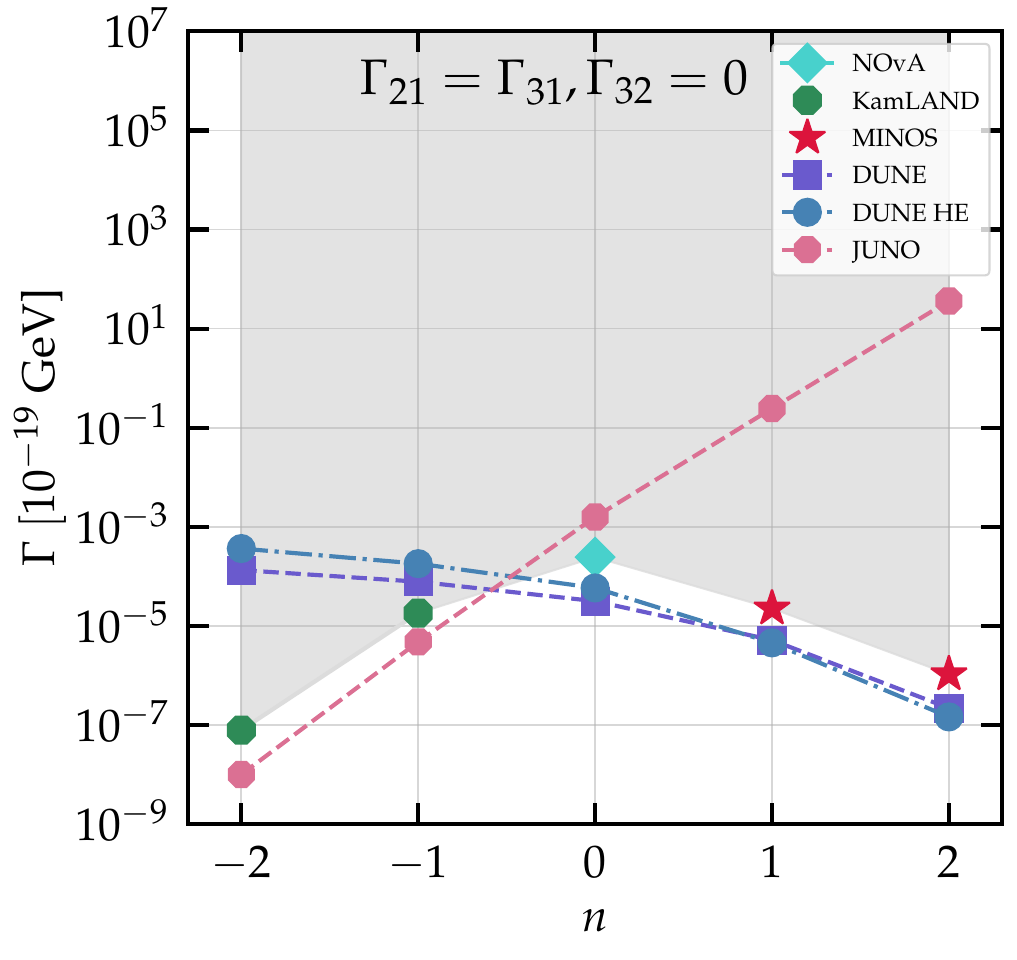}
\end{subfigure}
\caption{Current 90\% C.L. upper limits  (left) and future sensitivities for Model B ($\Gamma \equiv \Gamma_{21}=\Gamma_{31}$ and $\Gamma_{32}=0$), as a function of $n$ (see Eq.~\eqref{eq:gamma_E}).}
\label{fig:gam_21_31}
\end{figure}

\begin{figure}
\begin{subfigure}{0.49\textwidth}
    \includegraphics[width=\textwidth]{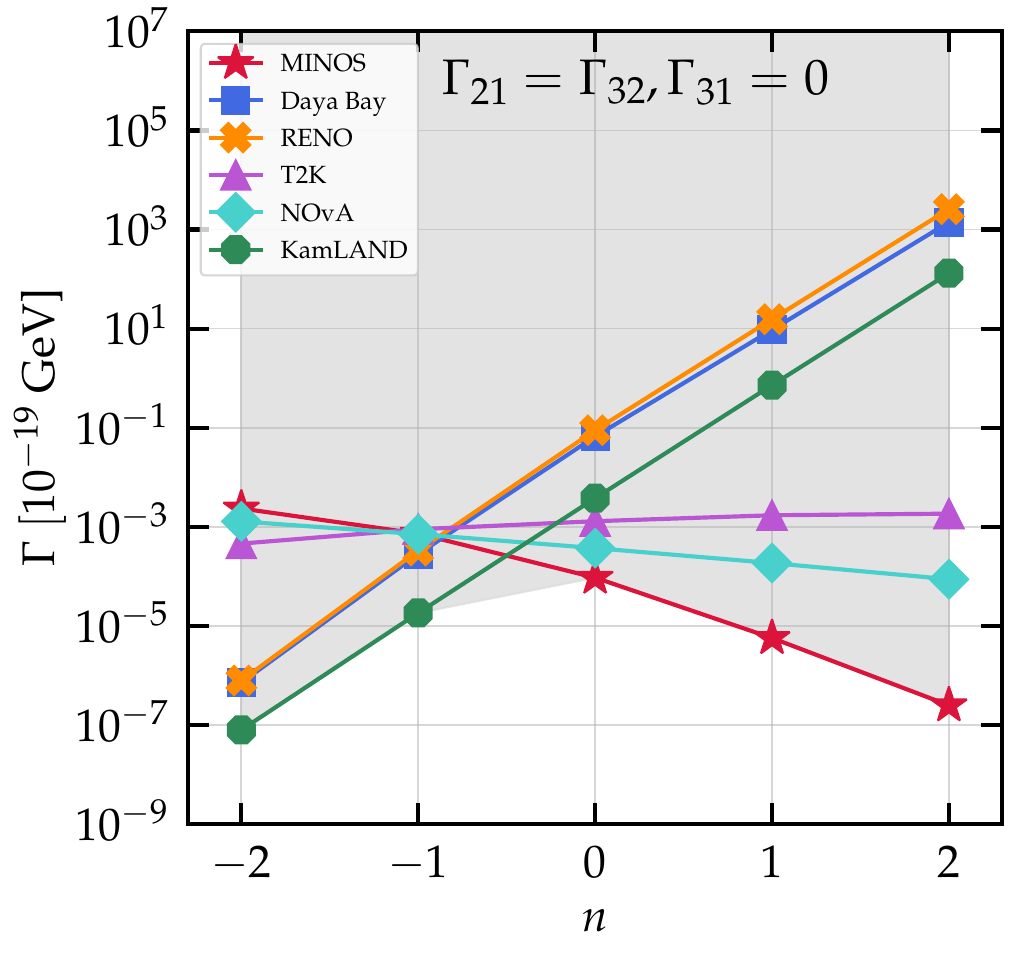}
\end{subfigure}
\hfill
\begin{subfigure}{0.49\textwidth}
    \includegraphics[width=\textwidth]{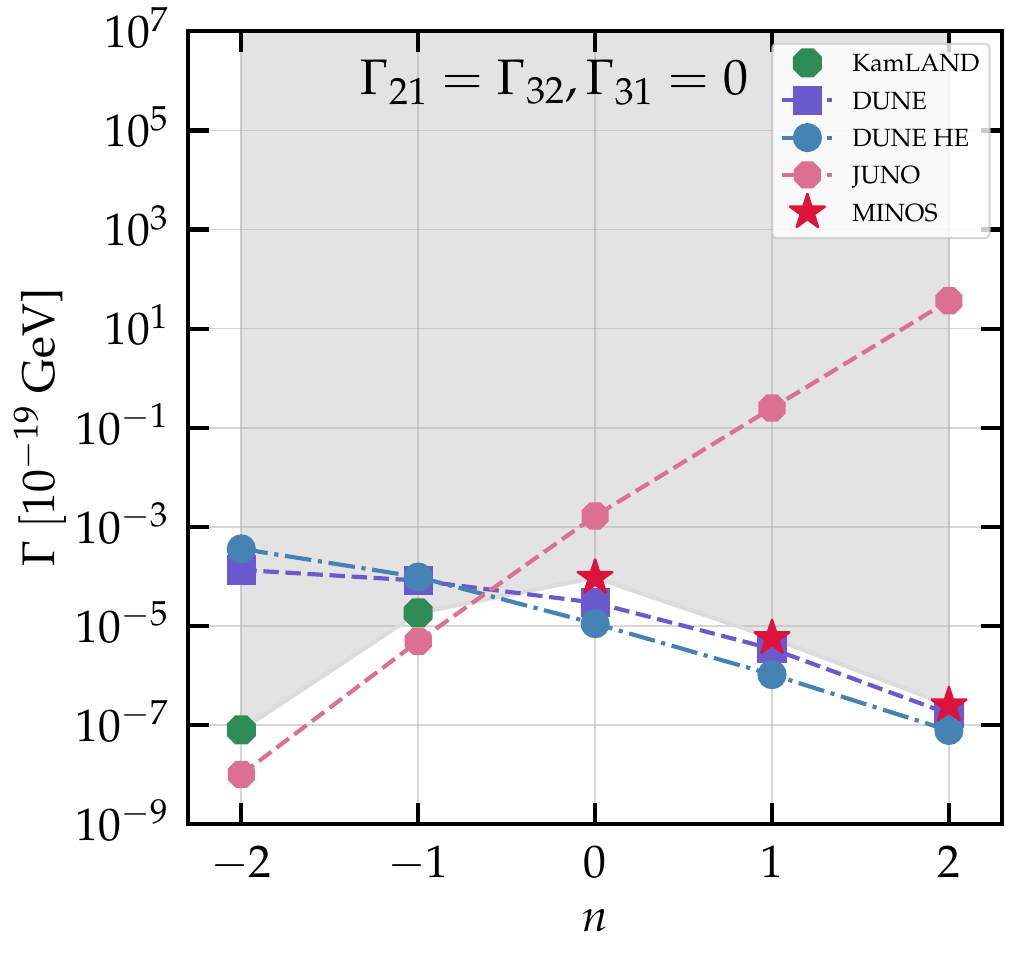}
\end{subfigure}
\caption{Current 90\% C.L. upper limits  (left) and future sensitivities for Model C ($\Gamma \equiv \Gamma_{21}=\Gamma_{32}$ and $\Gamma_{31}=0$), as a function of $n$ (see Eq.~\eqref{eq:gamma_E}).}
\label{fig:gam_21_32}
\end{figure}

\begin{figure}
\begin{subfigure}{0.49\textwidth}
    \includegraphics[width=\textwidth]{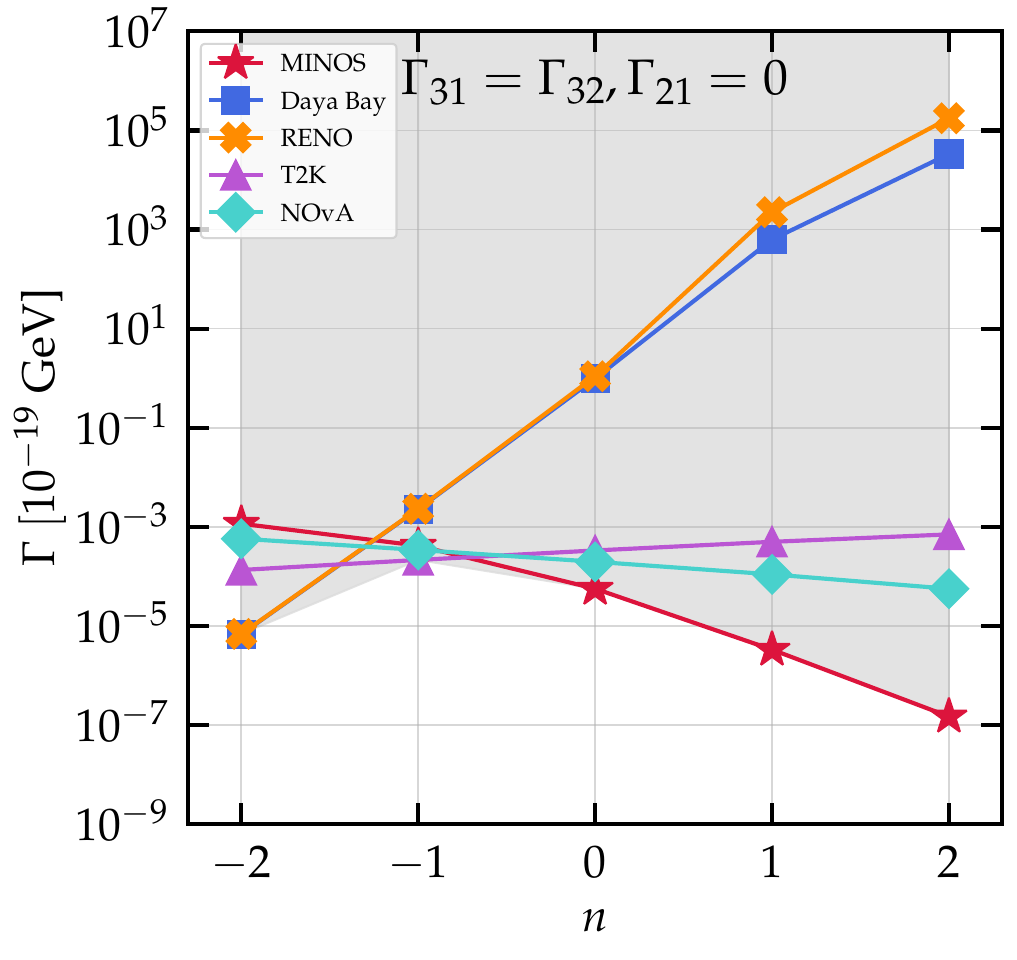}
\end{subfigure}
\hfill
\begin{subfigure}{0.49\textwidth}
    \includegraphics[width=\textwidth]{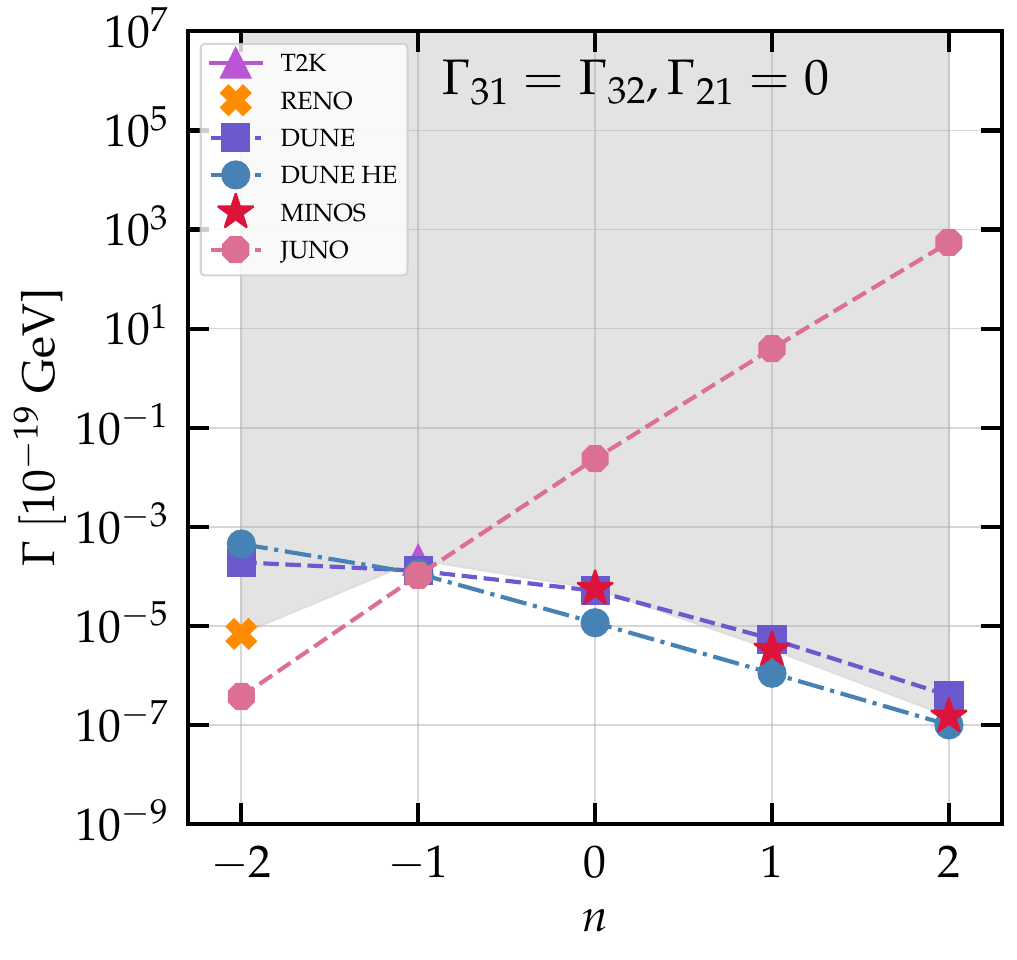}
\end{subfigure}
\caption{Current 90\% C.L. upper limits  (left) and future sensitivities for Model D ($\Gamma \equiv \Gamma_{31}=\Gamma_{32}$ and $\Gamma_{21}=0$), as a function of $n$ (see Eq.~\eqref{eq:gamma_E}).}
\label{fig:gam_31_32}
\end{figure}

\begin{figure}
\begin{subfigure}{0.49\textwidth}
    \includegraphics[width=\textwidth]{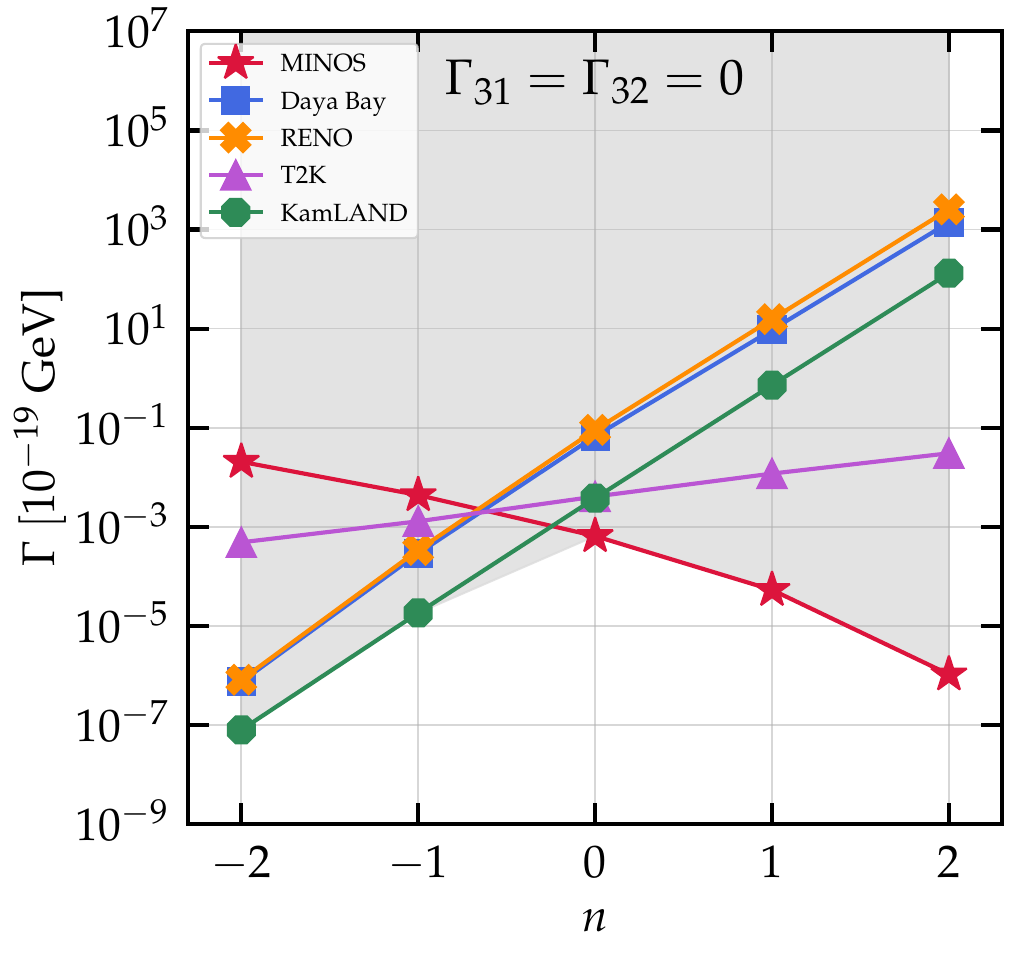}
\end{subfigure}
\hfill
\begin{subfigure}{0.49\textwidth}
    \includegraphics[width=\textwidth]{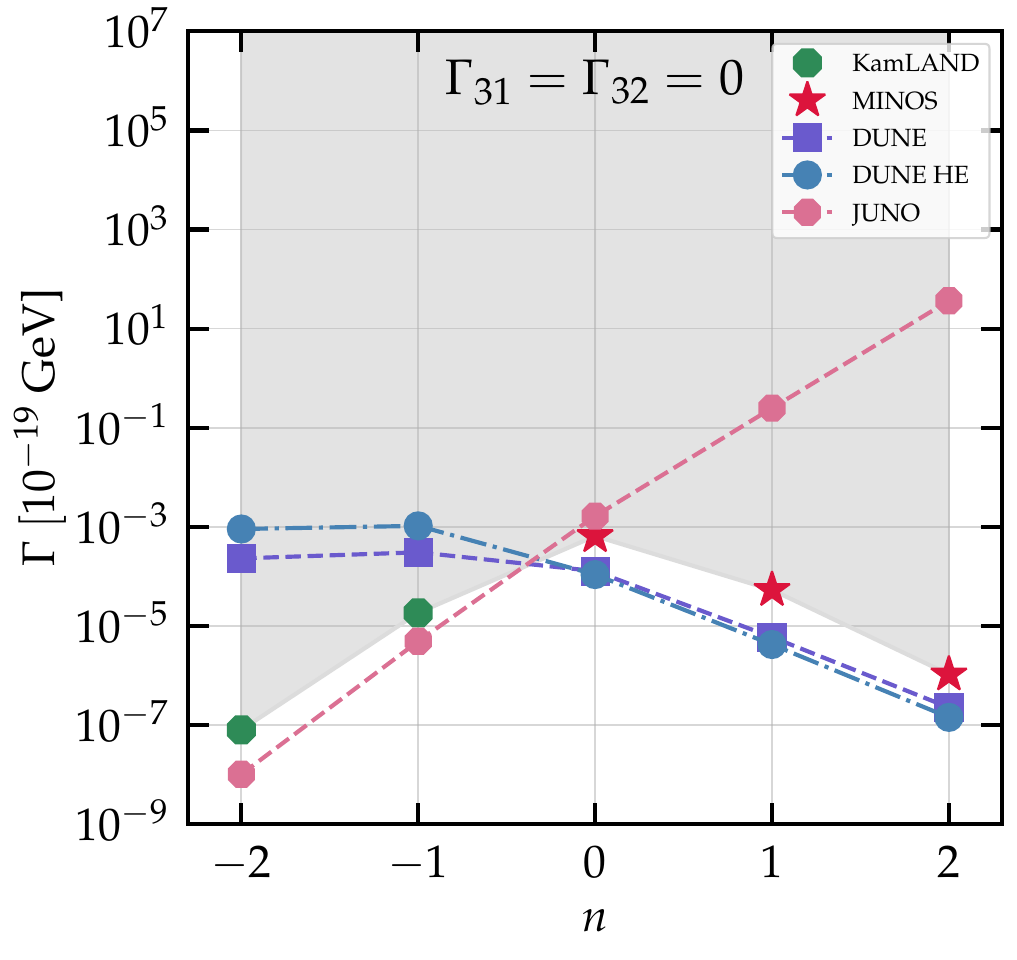}
\end{subfigure}
\caption{Current 90\% C.L. upper limits  (left) and future sensitivities for Model E ($\Gamma \equiv \Gamma_{21}$ and $\Gamma_{31}=\Gamma_{32}=0$), as a function of $n$ (see Eq.~\eqref{eq:gamma_E}).} 
\label{fig:gam_21}
\end{figure}

\begin{figure}
\begin{subfigure}{0.49\textwidth}
    \includegraphics[width=\textwidth]{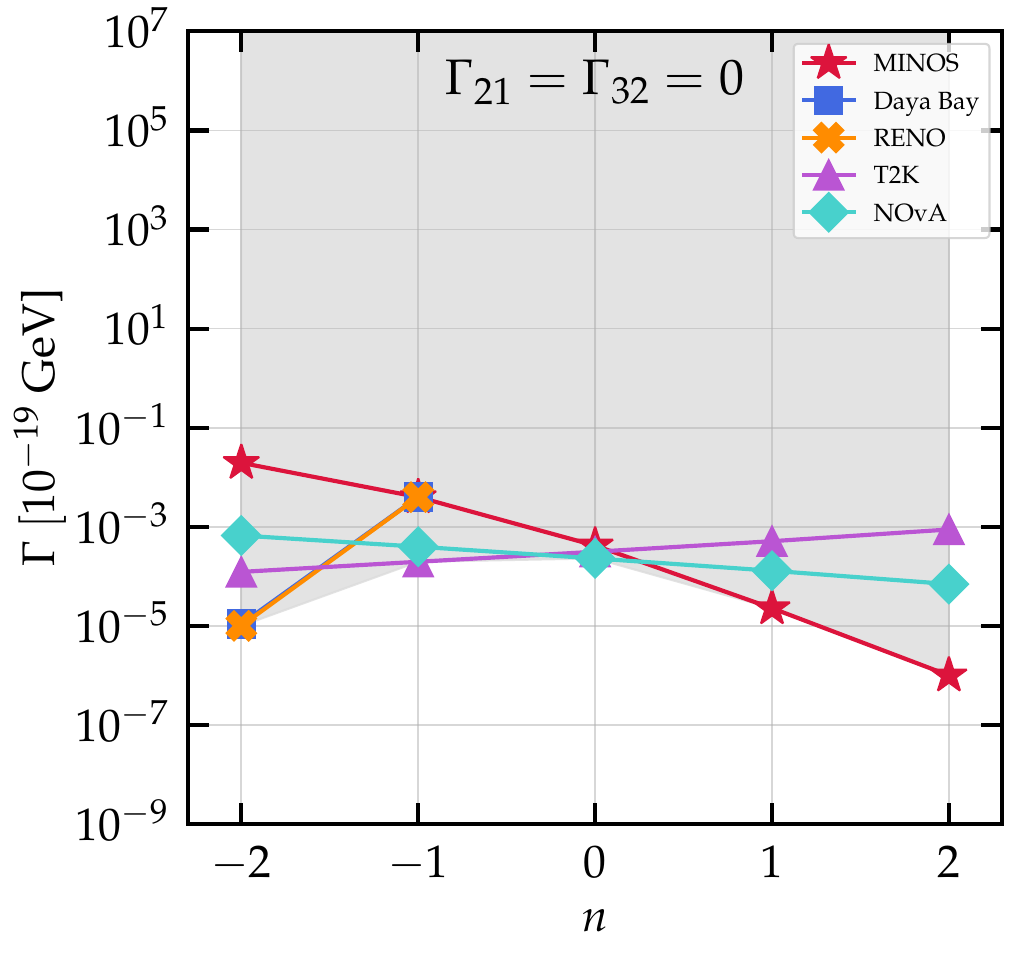}
\end{subfigure}
\hfill
\begin{subfigure}{0.49\textwidth}
    \includegraphics[width=\textwidth]{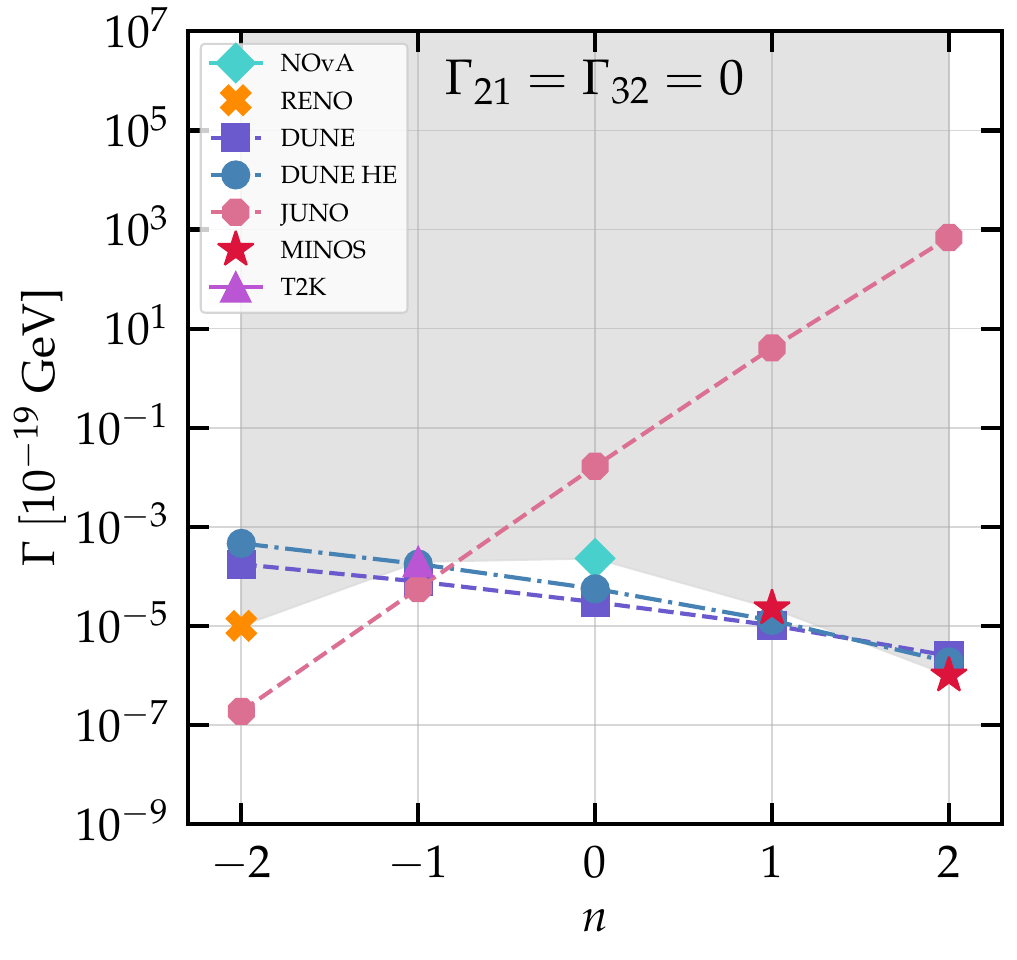}
\end{subfigure}
\caption{Current 90\% C.L. upper limits  (left) and future sensitivities for Model F ($\Gamma \equiv \Gamma_{31}$ and $\Gamma_{21}=\Gamma_{32}=0$), as a function of $n$ (see Eq.~\eqref{eq:gamma_E}).}    
\label{fig:gam_31}
\end{figure}

\begin{figure}
\centering
\begin{subfigure}{0.49\textwidth}
    \includegraphics[width=\textwidth]{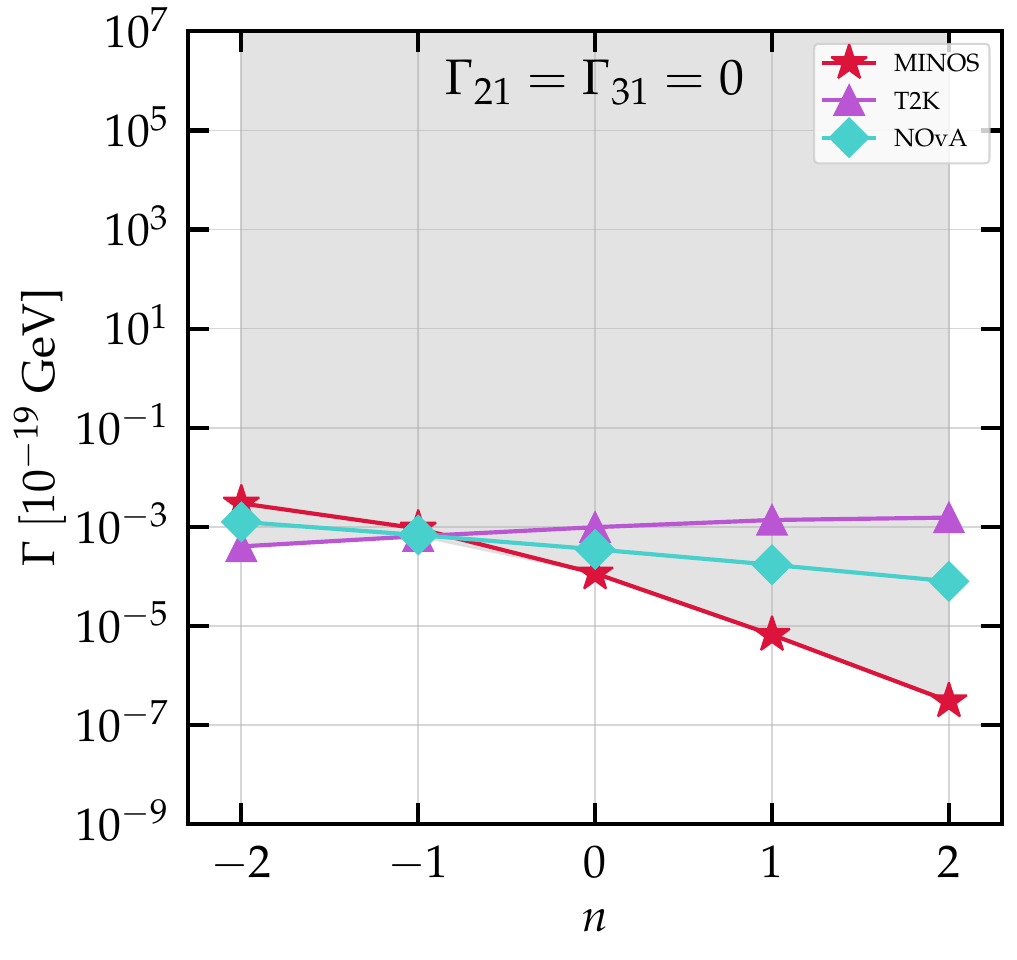}
\end{subfigure}
\hfill
\begin{subfigure}{0.49\textwidth}
    \includegraphics[width=\textwidth]{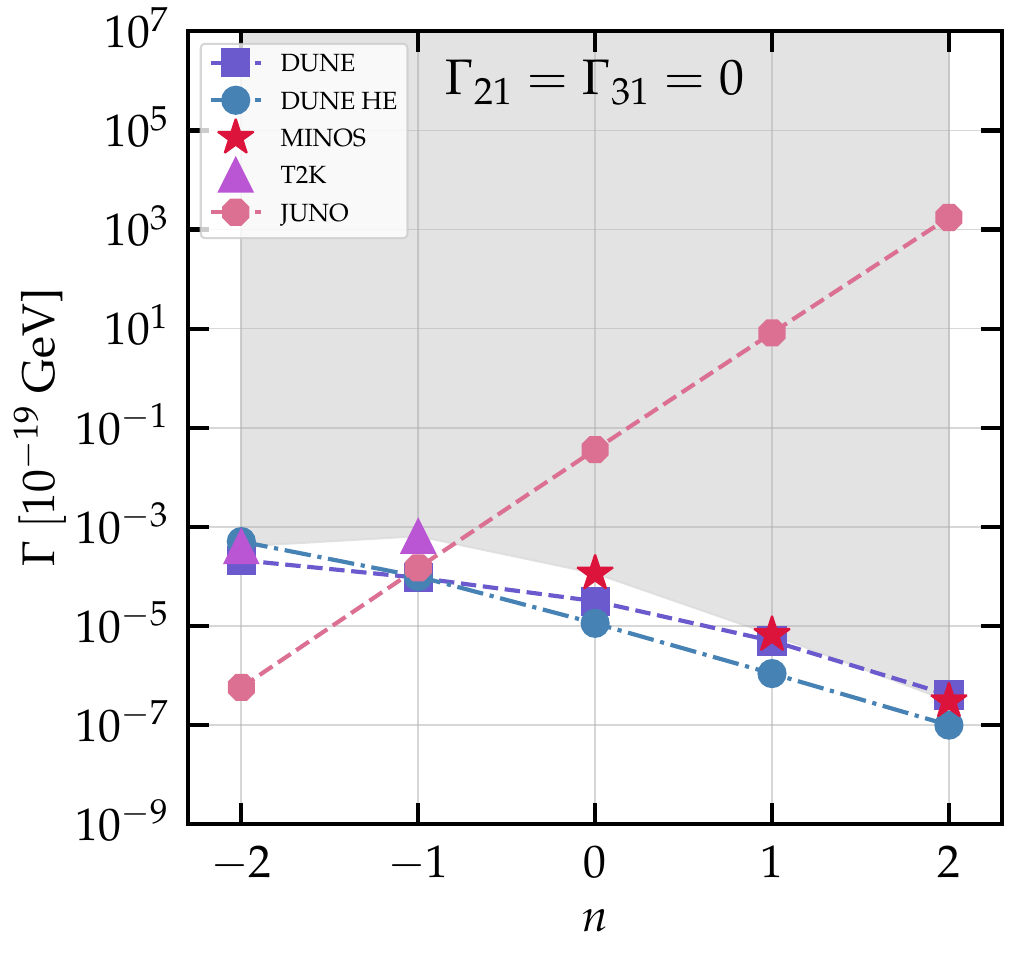}
\end{subfigure}
\caption{Current 90\% C.L. upper limits  (left) and future sensitivities for Model G ($\Gamma \equiv \Gamma_{32}$ and $\Gamma_{21}=\Gamma_{31}=0$), as a function of $n$ (see Eq.~\eqref{eq:gamma_E}).}    
\label{fig:gam_32}
\end{figure}

The results of our analyses for all the models under consideration are presented in Figs.~\ref{fig:all_gam}-\ref{fig:gam_32}. In the left panels we show the 90\% C.L.  upper limits that can be obtained with current data from the different experiments described in Sec.~\ref{sec:experiments}, for each $n$. In the right panels we show the sensitivities for JUNO and DUNE (using the standard and the high-energy flux configuration), always in comparison with the current strongest bound, obtained from the different experiments at each $n$ and represented by the grey-shaded region. 

In Fig.~\ref{fig:all_gam} we consider Model A ($\Gamma \equiv \Gamma_{21}=\Gamma_{31}=\Gamma_{32}$). As can be seen, for negative $n$ the strongest bound is obtained from the analysis of KamLAND data. For $n=0$ all accelerators produce similar results, while for positive $n$ the bound obtained from MINOS/MINOS+ is the strongest by far. The bounds obtained from the analysis of Daya Bay and RENO data are very similar and always about one order of magnitude weaker than the bound from KamLAND. Note that the slope of the  corresponding limits, as a function of $n$, is different for each experiment. This is due to the pivot point $E_0=1$~GeV chosen in Eq.~\eqref{eq:gamma_E}. Notice that the choice of a different reference value would change the bounds obtained in this work by a factor $\left(\frac{E_{0(\mathrm{old})}}{E_{0(\mathrm{new})}}\right)^n$. Reactors use energies much smaller and hence they are mostly sensitive to negative $n$. The energies relevant for T2K are close to $E_0$ and therefore the curve is nearly horizontal. NOvA and MINOS use larger energies and hence they obtain stronger bounds for positive $n$. This feature is particularly striking in the case of MINOS/MINOS+. In this case the overall bound is dominated by the analysis of KamLAND and MINOS data. In the right panel of Fig.~\ref{fig:all_gam} we compare the current best bound for each $n$ with the sensitivities for JUNO and DUNE/DUNE-HE. We find that JUNO will be able to improve the bound for negative $n$, while DUNE will improve the bound for $n\geq0$. Note that the sensitivity for DUNE using the high-energy beam can improve or worsen when comparing with the standard beam depending on $n$. Since the standard energy beam peaks at smaller energies, stronger bounds can be obtained for negative energy dependencies than with the high-energy beam.

The bounds and sensitivities for Models B and C (Figs.~\ref{fig:gam_21_31} and~\ref{fig:gam_21_32}) are very similar to the ones obtained for Model A, for $n<0$. As anticipated in Section~\ref{sec:osc_prob} we see that as long as $\Gamma_{21}$ and either $\Gamma_{31}$ or $\Gamma_{32}$ are active, the effect on the oscillation probability at reactors is similar to the case when all $\Gamma$ are switched on, and hence the results are nearly the same. For positive $n$ the bound is dominated by the analysis of data from accelerators. In this case we find that the sensitivity is better in Model C than in Model B, since Model C uses $\Gamma_{32}$ and Model B $\Gamma_{31}$, which has a smaller effect on the oscillation probability as we saw in Section~\ref{sec:osc_prob}.
The situation changes for model D ($\Gamma_{31}=\Gamma_{32}$ and $\Gamma_{21}=0$), shown in Fig.~\ref{fig:gam_31_32}. Since the fast oscillations due to $\Delta m_{31}^2$ and $\Delta m_{32}^2$ are averaged at KamLAND, KamLAND has no sensitivity at all to bound this model. For the other experiments the bounds on this model are again very similar to the ones obtained for Model A. Note that since JUNO will be able to resolve the fast oscillations, the JUNO sensitivity does not disappear as shown in the right panel of the figure.

For the remaining models only one parameter is active, $\Gamma_{21}$ (see Fig.~\ref{fig:gam_21}), $\Gamma_{31}$ (see Fig.~\ref{fig:gam_31}), and $\Gamma_{32}$ (see Fig.~\ref{fig:gam_32}).  
In the first case we find that all experiments except NOvA are capable of bounding $\Gamma_{21}$. It should be noted that for all experiments under consideration, except KamLAND and JUNO, the $\Delta m_{21}^2$-driven oscillation is not developed yet, and the effect of activating $\Gamma_{21}$ can be cancelled by modifying $\sin^2\theta_{12}$ accordingly. These experiments are capable of bounding $\Gamma_{21}$ because of the prior from the analysis of solar data~\cite{deSalas:2020pgw} imposed in our analyses. If we left $\sin^2\theta_{12}$ free in the analysis the bounds in this model would become weaker or would eventually disappear completely as in the case of NOvA. In the case of KamLAND and JUNO the main oscillation channel is directly affected and therefore we could always set a bound.
The other two cases (where we switch on only $\Gamma_{31}$ or $\Gamma_{32}$) are very similar. The differences in the bounds come from the fact that these models are either effecting the oscillation term which goes with $\Delta m_{31}^2$ or the one with $\Delta m_{32}^2$, which have slightly different amplitudes, as already discussed. In the case of the reactor neutrinos only in a few cases a bound can be obtained, because the data can be fit well using only one oscillating term with a modified amplitude. However,  experiments of next generation will be capable of probing all models for every energy dependence.

We summarize all bounds that we have obtained for each model in Table~\ref{tab:results}. As can be seen, the most constraining limit is often obtained in KamLAND and MINOS, although other experiments also contribute in a few cases.

\noindent
\begin{table}[!h]
\centering
\begin{threeparttable}
\resizebox{\linewidth}{!}{
\begin{tabular}{ l|c|c|c|c|c} 
\toprule
\textbf{Decoherence Model} & $\mathbf{n =-2}$ & $\mathbf{n =-1}$ &$\mathbf{n =0}$&$\mathbf{n =+1}$ &$\mathbf{n =+2}$ \\ 
\midrule
A: $\Gamma_{21}=\Gamma_{31}=\Gamma_{32}$ &$7.8\times10^{-27}$ (KL) &$ 1.8\times10^{-24}$ (KL) &$ 5.1\times10^{-24}$ (M)&$ 3.0\times10^{-25}$ (M) &$ 1.3\times10^{-26}$ (M) \\
B: $\Gamma_{21}=\Gamma_{31}$, $\Gamma_{32}= 0$ &$7.9\times10^{-27}$ (KL) &$ 1.8\times10^{-24}$ (KL) &$ 2.4\times10^{-23}$ (N)&$ 2.3\times10^{-24}$ (M) &$ 1.0\times10^{-25}$ (M) \\
C: $\Gamma_{21}=\Gamma_{32}$, $\Gamma_{31}= 0$ &$7.9\times10^{-27}$ (KL) &$ 1.8\times10^{-24}$ (KL) &$ 9.4\times10^{-24}$ (M)&$ 5.7\times10^{-25}$ (M) &$ 2.5\times10^{-26}$ (M)  \\
D: $\Gamma_{31}=\Gamma_{32}$, $\Gamma_{21}= 0$&
$6.9\times10^{-25}$ (R) &$ 2.1\times10^{-23}$ (T2K) &$ 5.6\times10^{-24}$ (M)&$ 3.3\times10^{-25}$ (M) &$ 1.5\times10^{-26}$ (M)\\
E: $\Gamma_{21}$, $\Gamma_{31}=\Gamma_{32}= 0$ &$7.9\times10^{-27}$ (KL) &$ 1.8\times10^{-24}$ (KL) &$ 3.2\times10^{-23}$ (M)&$ 2.2\times10^{-24}$ (M) &$ 1.0\times10^{-25}$ (M)\\
F: $\Gamma_{31}$, $\Gamma_{21}=\Gamma_{32}= 0$ &$1.0\times10^{-24}$ (R) &$ 1.9\times10^{-23}$ (T2K) &$ 2.3\times10^{-23}$ (N)&$ 2.2\times10^{-24}$ (M) &$ 1.0\times10^{-25}$ (M)\\
G: $\Gamma_{32}$, $\Gamma_{21}=\Gamma_{31}= 0$ &$4.0\times10^{-23}$ (T2K) &$ 6.5\times10^{-23}$ (T2K) &$ 1.1\times10^{-23}$ (M)&$ 6.6\times10^{-25}$ (M) &$ 3.0\times10^{-26}$ (M) \\
\hline
\bottomrule
\end{tabular}}
\caption{Summary of results: each column shows the most constraining upper limit on $\Gamma_{ij}$, in GeV, for each model (A to G) and value of $n$. We also clarify, within parenthesis, which experiment sets the bound (KL = KamLAND, R = RENO, M = MINOS/MINOS+, N = NOvA).  }
\label{tab:results}
\end{threeparttable}
\end{table}

\section{Conclusions}
\label{sec:concl}
Working in the open quantum system framework, we have analyzed a set of neutrino oscillation experiments and we have obtained bounds on quantum decoherence for several phenomenological models.
We parametrize the energy dependence of such quantum decoherence effects as $\propto E^{n}$, and consider both positive exponents $n>0$, an energy-independent scenario, $n=0$, and negative exponents $n<0$.
Our analysis of MINOS/MINOS+ and T2K data updates the results that were obtained in Ref.~\cite{Gomes:2020muc}, while providing the first analysis of NOvA data in this context. The data of KamLAND had already been analyzed in Ref.~\cite{BalieiroGomes:2016ykp} using $n=-1,0,1$. For these cases we find good agreement on the bounds with the authors of Ref.~\cite{BalieiroGomes:2016ykp} and in addition we provide the bounds for $n=\pm 2$. Our analyses provide the strongest bounds to date for negative values of $n$. It should be further noted that very strong bounds were obtained from a combined analyses of solar and KamLAND data in Ref.~\cite{Fogli:2007tx}. However, these limits are model-dependent and in particular do not apply to a scenario in which only decoherence effects are included (as it is always the case in this paper), as was already argued in Refs.~\cite{BalieiroGomes:2016ykp,Guzzo:2014jbp}. We find that the bounds for negative $n$ will be improved by about one or two orders of magnitude, depending on the model under consideration, at JUNO. Regarding $n\geq0$ the bounds obtained from MINOS/MINOS+, T2K, and NOvA, can be improved using DUNE and DUNE-HE. The capabilities of DUNE where already discussed in Ref.~\cite{BalieiroGomes:2018gtd} focusing only on $n=0$. For completeness, we note that very strong bounds on quantum decoherence can also obtained from atmospheric neutrinos~\cite{Coloma:2018idr}.

As a final note, in this paper we have performed analyses of each experiment individually. We only imposed a prior on $\sin^2\theta_{12}$ from solar neutrino measurements, and after finding that KamLAND's $\Delta m_{21}^2$ measurement is robust under decoherence we added another prior on this parameter in the analysis of the remaining experiments. The measurements of the standard parameters at the other experiments might not be robust anymore. We have found, for example, that the measurement of $\sin^2\theta_{13}$ at Daya Bay and RENO is affected, and therefore this parameter must be left to vary freely in the analyses of the remaining experiments. A truly combined analysis might improve a bit the bounds that we have obtained here. The measurements of the standard parameters might be more precise from a combination of experiments and might therefore break the degeneracies with the $\Gamma_{ij}$ leading to slightly stronger bounds. Since the improvement from such a complicated analysis is not expected to be very large, we decided to focus on the experiments individually.

\section*{Acknowledgments}
This work has been supported by a STSM Grant from the COST Action “Quantum gravity phenomenology in the multi-messenger approach”, CA18108.
C.A.T. is very thankful for the hospitality at the Niels Bohr Institute in Copenhagen where part of this work has been performed.
V.D.R. is grateful for the kind hospitality received at the INFN and Physics Department of the University of Torino, where part of the work was done.
V.D.R. acknowledges financial support by the CIDEXG/2022/20 grant (project ``D'AMAGAT'') funded by
Generalitat Valenciana and by the Spanish grant PID2020-113775GB-I00 (MCIN/AEI/10.13039/501100011033).
C.G. and C.A.T. are supported by the research grant ``The Dark Universe: A Synergic Multimessenger Approach'' number 2017X7X85K under the program ``PRIN 2017'' funded by the Italian Ministero dell'Istruzione, Universit\`a e della Ricerca (MIUR). C.A.T. also acknowledges support from {\sl Departments of Excellence} grant awarded by MIUR and the research grant {\sl TAsP (Theoretical Astroparticle Physics)} funded by Istituto Nazionale di Fisica Nucleare (INFN).
T.S. is supported by research grant CF19-0652 funded by Carlsbergfondet.

\noindent

\appendix


\begin{thebibliography}{10}

\bibitem{Kajita:2016cak}
T.~Kajita, ``{Nobel Lecture: Discovery of atmospheric neutrino oscillations},''
  \href{http://dx.doi.org/10.1103/RevModPhys.88.030501}{{\em Rev. Mod. Phys.}
  {\bfseries 88} (2016) 030501}.

\bibitem{McDonald:2016ixn}
A.~B. McDonald, ``{Nobel Lecture: The Sudbury Neutrino Observatory: Observation
  of flavor change for solar neutrinos},''
  \href{http://dx.doi.org/10.1103/RevModPhys.88.030502}{{\em Rev. Mod. Phys.}
  {\bfseries 88} (2016) 030502}.

\bibitem{deSalas:2020pgw}
P.~F. de~Salas, D.~V. Forero, S.~Gariazzo, P.~Mart\'\i{}nez-Mirav\'e, O.~Mena,
  C.~A. Ternes, M.~T\'ortola, and J.~W.~F. Valle, ``{2020 global reassessment
  of the neutrino oscillation picture},''
  \href{http://dx.doi.org/10.1007/JHEP02(2021)071}{{\em JHEP} {\bfseries 02}
  (2021) 071}, \href{http://arxiv.org/abs/2006.11237}{{\ttfamily
  arXiv:2006.11237 [hep-ph]}}.

\bibitem{Esteban:2020cvm}
I.~Esteban, M.~C. Gonzalez-Garcia, M.~Maltoni, T.~Schwetz, and A.~Zhou, ``{The
  fate of hints: updated global analysis of three-flavor neutrino
  oscillations},'' \href{http://dx.doi.org/10.1007/JHEP09(2020)178}{{\em JHEP}
  {\bfseries 09} (2020) 178}, \href{http://arxiv.org/abs/2007.14792}{{\ttfamily
  arXiv:2007.14792 [hep-ph]}}.

\bibitem{Capozzi:2021fjo}
F.~Capozzi, E.~Di~Valentino, E.~Lisi, A.~Marrone, A.~Melchiorri, and
  A.~Palazzo, ``{Unfinished fabric of the three neutrino paradigm},''
  \href{http://dx.doi.org/10.1103/PhysRevD.104.083031}{{\em Phys. Rev. D}
  {\bfseries 104} no.~8, (2021) 083031},
  \href{http://arxiv.org/abs/2107.00532}{{\ttfamily arXiv:2107.00532
  [hep-ph]}}.

\bibitem{Kiers:1995zj}
K.~Kiers, S.~Nussinov, and N.~Weiss, ``{Coherence effects in neutrino
  oscillations},'' \href{http://dx.doi.org/10.1103/PhysRevD.53.537}{{\em Phys.
  Rev. D} {\bfseries 53} (1996) 537--547},
  \href{http://arxiv.org/abs/hep-ph/9506271}{{\ttfamily arXiv:hep-ph/9506271}}.

\bibitem{Ohlsson:2000mj}
T.~Ohlsson, ``{Equivalence between neutrino oscillations and neutrino
  decoherence},'' \href{http://dx.doi.org/10.1016/S0370-2693(01)00178-2}{{\em
  Phys. Lett. B} {\bfseries 502} (2001) 159--166},
  \href{http://arxiv.org/abs/hep-ph/0012272}{{\ttfamily arXiv:hep-ph/0012272}}.

\bibitem{Beuthe:2001rc}
M.~Beuthe, ``{Oscillations of neutrinos and mesons in quantum field theory},''
  \href{http://dx.doi.org/10.1016/S0370-1573(02)00538-0}{{\em Phys. Rept.}
  {\bfseries 375} (2003) 105--218},
  \href{http://arxiv.org/abs/hep-ph/0109119}{{\ttfamily arXiv:hep-ph/0109119}}.

\bibitem{Beuthe:2002ej}
M.~Beuthe, ``{Towards a unique formula for neutrino oscillations in vacuum},''
  \href{http://dx.doi.org/10.1103/PhysRevD.66.013003}{{\em Phys. Rev. D}
  {\bfseries 66} (2002) 013003},
  \href{http://arxiv.org/abs/hep-ph/0202068}{{\ttfamily arXiv:hep-ph/0202068}}.

\bibitem{Giunti:2003ax}
C.~Giunti, ``{Coherence and wave packets in neutrino oscillations},''
  \href{http://dx.doi.org/10.1023/B:FOPL.0000019651.53280.31}{{\em Found. Phys.
  Lett.} {\bfseries 17} (2004) 103--124},
  \href{http://arxiv.org/abs/hep-ph/0302026}{{\ttfamily arXiv:hep-ph/0302026}}.

\bibitem{Blennow:2005yk}
M.~Blennow, T.~Ohlsson, and W.~Winter, ``{Damping signatures in future neutrino
  oscillation experiments},''
  \href{http://dx.doi.org/10.1088/1126-6708/2005/06/049}{{\em JHEP} {\bfseries
  06} (2005) 049}, \href{http://arxiv.org/abs/hep-ph/0502147}{{\ttfamily
  arXiv:hep-ph/0502147}}.

\bibitem{Farzan:2008eg}
Y.~Farzan and A.~Y. Smirnov, ``{Coherence and oscillations of cosmic
  neutrinos},'' \href{http://dx.doi.org/10.1016/j.nuclphysb.2008.07.028}{{\em
  Nucl. Phys. B} {\bfseries 805} (2008) 356--376},
  \href{http://arxiv.org/abs/0803.0495}{{\ttfamily arXiv:0803.0495 [hep-ph]}}.

\bibitem{Kayser:2010pr}
B.~Kayser and J.~Kopp, ``{Testing the Wave Packet Approach to Neutrino
  Oscillations in Future Experiments},''
  \href{http://arxiv.org/abs/1005.4081}{{\ttfamily arXiv:1005.4081 [hep-ph]}}.

\bibitem{Jones:2014sfa}
B.~J.~P. Jones, ``{Dynamical pion collapse and the coherence of conventional
  neutrino beams},'' \href{http://dx.doi.org/10.1103/PhysRevD.91.053002}{{\em
  Phys. Rev. D} {\bfseries 91} no.~5, (2015) 053002},
  \href{http://arxiv.org/abs/1412.2264}{{\ttfamily arXiv:1412.2264 [hep-ph]}}.

\bibitem{Akhmedov:2019iyt}
E.~Akhmedov, ``{Quantum mechanics aspects and subtleties of neutrino
  oscillations},'' in {\em {International Conference on History of the
  Neutrino}: {1930-2018}}.
\newblock 1, 2019.
\newblock \href{http://arxiv.org/abs/1901.05232}{{\ttfamily arXiv:1901.05232
  [hep-ph]}}.

\bibitem{Grimus:2019hlq}
W.~Grimus, ``{Revisiting the quantum field theory of neutrino oscillations in
  vacuum},'' \href{http://dx.doi.org/10.1088/1361-6471/ab716f}{{\em J. Phys. G}
  {\bfseries 47} no.~8, (2020) 085004},
  \href{http://arxiv.org/abs/1910.13776}{{\ttfamily arXiv:1910.13776
  [hep-ph]}}.

\bibitem{Naumov:2020yyv}
D.~V. Naumov and V.~A. Naumov, ``{Quantum Field Theory of Neutrino
  Oscillations},'' \href{http://dx.doi.org/10.1134/S1063779620010050}{{\em
  Phys. Part. Nucl.} {\bfseries 51} no.~1, (2020) 1--106}.

\bibitem{Akhmedov:2022bjs}
E.~Akhmedov and A.~Y. Smirnov, ``{Damping of neutrino oscillations, decoherence
  and the lengths of neutrino wave packets},''
  \href{http://dx.doi.org/10.1007/JHEP11(2022)082}{{\em JHEP} {\bfseries 11}
  (2022) 082}, \href{http://arxiv.org/abs/2208.03736}{{\ttfamily
  arXiv:2208.03736 [hep-ph]}}.

\bibitem{Krueger:2023skk}
R.~Krueger and T.~Schwetz, ``{Decoherence effects in reactor and Gallium
  neutrino oscillation experiments -- a QFT approach},''
  \href{http://arxiv.org/abs/2303.15524}{{\ttfamily arXiv:2303.15524
  [hep-ph]}}.

\bibitem{Giunti:1991sx}
C.~Giunti, C.~W. Kim, and U.~W. Lee, ``{Coherence of neutrino oscillations in
  vacuum and matter in the wave packet treatment},''
  \href{http://dx.doi.org/10.1016/0370-2693(92)90308-Q}{{\em Phys. Lett. B}
  {\bfseries 274} (1992) 87--94}.

\bibitem{Naumov:2013uia}
D.~V. Naumov, ``{On the Theory of Wave Packets},''
  \href{http://dx.doi.org/10.1134/S1547477113070145}{{\em Phys. Part. Nucl.
  Lett.} {\bfseries 10} (2013) 642--650},
  \href{http://arxiv.org/abs/1309.1717}{{\ttfamily arXiv:1309.1717
  [quant-ph]}}.

\bibitem{deGouvea:2020hfl}
A.~de~Gouvêa, V.~De~Romeri, and C.~A. Ternes, ``{Probing neutrino quantum
  decoherence at reactor experiments},''
  \href{http://dx.doi.org/10.1007/JHEP08(2020)049}{{\em JHEP} {\bfseries 08}
  (2020) 018}, \href{http://arxiv.org/abs/2005.03022}{{\ttfamily
  arXiv:2005.03022 [hep-ph]}}.

\bibitem{deGouvea:2021uvg}
A.~de~Gouv\^ea, V.~De~Romeri, and C.~A. Ternes, ``{Combined analysis of
  neutrino decoherence at reactor experiments},''
  \href{http://dx.doi.org/10.1007/JHEP06(2021)042}{{\em JHEP} {\bfseries 06}
  (2021) 042}, \href{http://arxiv.org/abs/2104.05806}{{\ttfamily
  arXiv:2104.05806 [hep-ph]}}.

\bibitem{Breuer:2002pc}
H.~P. Breuer and F.~Petruccione, {\em {The theory of open quantum systems}}.
\newblock 2002.

\bibitem{Gago:2000qc}
A.~M. Gago, E.~M. Santos, W.~J.~C. Teves, and R.~Zukanovich~Funchal, ``{Quantum
  dissipative effects and neutrinos: Current constraints and future
  perspectives},'' \href{http://dx.doi.org/10.1103/PhysRevD.63.073001}{{\em
  Phys. Rev. D} {\bfseries 63} (2001) 073001},
  \href{http://arxiv.org/abs/hep-ph/0009222}{{\ttfamily arXiv:hep-ph/0009222}}.

\bibitem{Benatti:2000ph}
F.~Benatti and R.~Floreanini, ``{Open system approach to neutrino
  oscillations},'' \href{http://dx.doi.org/10.1088/1126-6708/2000/02/032}{{\em
  JHEP} {\bfseries 02} (2000) 032},
  \href{http://arxiv.org/abs/hep-ph/0002221}{{\ttfamily arXiv:hep-ph/0002221}}.

\bibitem{Lisi:2000zt}
E.~Lisi, A.~Marrone, and D.~Montanino, ``{Probing possible decoherence effects
  in atmospheric neutrino oscillations},''
  \href{http://dx.doi.org/10.1103/PhysRevLett.85.1166}{{\em Phys. Rev. Lett.}
  {\bfseries 85} (2000) 1166--1169},
  \href{http://arxiv.org/abs/hep-ph/0002053}{{\ttfamily arXiv:hep-ph/0002053}}.

\bibitem{Benatti:2001fa}
F.~Benatti and R.~Floreanini, ``{Massless neutrino oscillations},''
  \href{http://dx.doi.org/10.1103/PhysRevD.64.085015}{{\em Phys. Rev. D}
  {\bfseries 64} (2001) 085015},
  \href{http://arxiv.org/abs/hep-ph/0105303}{{\ttfamily arXiv:hep-ph/0105303}}.

\bibitem{Morgan:2004vv}
D.~Morgan, E.~Winstanley, J.~Brunner, and L.~F. Thompson, ``{Probing quantum
  decoherence in atmospheric neutrino oscillations with a neutrino
  telescope},''
  \href{http://dx.doi.org/10.1016/j.astropartphys.2006.03.001}{{\em Astropart.
  Phys.} {\bfseries 25} (2006) 311--327},
  \href{http://arxiv.org/abs/astro-ph/0412618}{{\ttfamily
  arXiv:astro-ph/0412618}}.

\bibitem{Anchordoqui:2005gj}
L.~A. Anchordoqui, H.~Goldberg, M.~C. Gonzalez-Garcia, F.~Halzen, D.~Hooper,
  S.~Sarkar, and T.~J. Weiler, ``{Probing Planck scale physics with IceCube},''
  \href{http://dx.doi.org/10.1103/PhysRevD.72.065019}{{\em Phys. Rev. D}
  {\bfseries 72} (2005) 065019},
  \href{http://arxiv.org/abs/hep-ph/0506168}{{\ttfamily arXiv:hep-ph/0506168}}.

\bibitem{Fogli:2007tx}
G.~L. Fogli, E.~Lisi, A.~Marrone, D.~Montanino, and A.~Palazzo, ``{Probing
  non-standard decoherence effects with solar and KamLAND neutrinos},''
  \href{http://dx.doi.org/10.1103/PhysRevD.76.033006}{{\em Phys. Rev. D}
  {\bfseries 76} (2007) 033006},
  \href{http://arxiv.org/abs/0704.2568}{{\ttfamily arXiv:0704.2568 [hep-ph]}}.

\bibitem{Farzan:2008zv}
Y.~Farzan, T.~Schwetz, and A.~Y. Smirnov, ``{Reconciling results of LSND,
  MiniBooNE and other experiments with soft decoherence},''
  \href{http://dx.doi.org/10.1088/1126-6708/2008/07/067}{{\em JHEP} {\bfseries
  07} (2008) 067}, \href{http://arxiv.org/abs/0805.2098}{{\ttfamily
  arXiv:0805.2098 [hep-ph]}}.

\bibitem{Oliveira:2013nua}
R.~L.~N. Oliveira and M.~M. Guzzo, ``{Dissipation and $\theta_{13}$ in neutrino
  oscillations},'' \href{http://dx.doi.org/10.1140/epjc/s10052-013-2434-6}{{\em
  Eur. Phys. J. C} {\bfseries 73} (2013) 2434}.

\bibitem{Oliveira:2016asf}
R.~L.~N. Oliveira, ``{Dissipative Effect in Long Baseline Neutrino
  Experiments},'' \href{http://dx.doi.org/10.1140/epjc/s10052-016-4253-z}{{\em
  Eur. Phys. J. C} {\bfseries 76} no.~7, (2016) 417},
  \href{http://arxiv.org/abs/1603.08065}{{\ttfamily arXiv:1603.08065
  [hep-ph]}}.

\bibitem{BalieiroGomes:2016ykp}
G.~Balieiro~Gomes, M.~M. Guzzo, P.~C. de~Holanda, and R.~L.~N. Oliveira,
  ``{Parameter Limits for Neutrino Oscillation with Decoherence in KamLAND},''
  \href{http://dx.doi.org/10.1103/PhysRevD.95.113005}{{\em Phys. Rev. D}
  {\bfseries 95} no.~11, (2017) 113005},
  \href{http://arxiv.org/abs/1603.04126}{{\ttfamily arXiv:1603.04126
  [hep-ph]}}.

\bibitem{Coelho:2017zes}
J.~A.~B. Coelho, W.~A. Mann, and S.~S. Bashar, ``{Nonmaximal $\theta_{23}$
  mixing at NOvA from neutrino decoherence},''
  \href{http://dx.doi.org/10.1103/PhysRevLett.118.221801}{{\em Phys. Rev.
  Lett.} {\bfseries 118} no.~22, (2017) 221801},
  \href{http://arxiv.org/abs/1702.04738}{{\ttfamily arXiv:1702.04738
  [hep-ph]}}.

\bibitem{Coelho:2017byq}
J.~A.~B. Coelho and W.~A. Mann, ``{Decoherence, matter effect, and neutrino
  hierarchy signature in long baseline experiments},''
  \href{http://dx.doi.org/10.1103/PhysRevD.96.093009}{{\em Phys. Rev. D}
  {\bfseries 96} no.~9, (2017) 093009},
  \href{http://arxiv.org/abs/1708.05495}{{\ttfamily arXiv:1708.05495
  [hep-ph]}}.

\bibitem{Carpio:2017nui}
J.~Carpio, E.~Massoni, and A.~M. Gago, ``{Revisiting quantum decoherence for
  neutrino oscillations in matter with constant density},''
  \href{http://dx.doi.org/10.1103/PhysRevD.97.115017}{{\em Phys. Rev. D}
  {\bfseries 97} no.~11, (2018) 115017},
  \href{http://arxiv.org/abs/1711.03680}{{\ttfamily arXiv:1711.03680
  [hep-ph]}}.

\bibitem{Coloma:2018idr}
P.~Coloma, J.~Lopez-Pavon, I.~Martinez-Soler, and H.~Nunokawa, ``{Decoherence
  in Neutrino Propagation Through Matter, and Bounds from IceCube/DeepCore},''
  \href{http://dx.doi.org/10.1140/epjc/s10052-018-6092-6}{{\em Eur. Phys. J. C}
  {\bfseries 78} no.~8, (2018) 614},
  \href{http://arxiv.org/abs/1803.04438}{{\ttfamily arXiv:1803.04438
  [hep-ph]}}.

\bibitem{Carpio:2018gum}
J.~A. Carpio, E.~Massoni, and A.~M. Gago, ``{Testing quantum decoherence at
  DUNE},'' \href{http://dx.doi.org/10.1103/PhysRevD.100.015035}{{\em Phys. Rev.
  D} {\bfseries 100} no.~1, (2019) 015035},
  \href{http://arxiv.org/abs/1811.07923}{{\ttfamily arXiv:1811.07923
  [hep-ph]}}.

\bibitem{Carrasco:2018sca}
J.~C. Carrasco, F.~N. D\'\i{}az, and A.~M. Gago, ``{Probing CPT breaking
  induced by quantum decoherence at DUNE},''
  \href{http://dx.doi.org/10.1103/PhysRevD.99.075022}{{\em Phys. Rev. D}
  {\bfseries 99} no.~7, (2019) 075022},
  \href{http://arxiv.org/abs/1811.04982}{{\ttfamily arXiv:1811.04982
  [hep-ph]}}.

\bibitem{Buoninfante:2020iyr}
L.~Buoninfante, A.~Capolupo, S.~M. Giampaolo, and G.~Lambiase, ``{Revealing
  neutrino nature and $CPT$ violation with decoherence effects},''
  \href{http://dx.doi.org/10.1140/epjc/s10052-020-08549-9}{{\em Eur. Phys. J.
  C} {\bfseries 80} no.~11, (2020) 1009},
  \href{http://arxiv.org/abs/2001.07580}{{\ttfamily arXiv:2001.07580
  [hep-ph]}}.

\bibitem{Gomes:2020muc}
A.~L.~G. Gomes, R.~A. Gomes, and O.~L.~G. Peres, ``{Quantum decoherence and
  relaxation in neutrinos using long-baseline data},''
  \href{http://arxiv.org/abs/2001.09250}{{\ttfamily arXiv:2001.09250
  [hep-ph]}}.

\bibitem{Ohlsson:2020gxx}
T.~Ohlsson and S.~Zhou, ``{Density Matrix Formalism for PT-Symmetric
  Non-Hermitian Hamiltonians with the Lindblad Equation},''
  \href{http://dx.doi.org/10.1103/PhysRevA.103.022218}{{\em Phys. Rev. A}
  {\bfseries 103} no.~2, (2021) 022218},
  \href{http://arxiv.org/abs/2006.02445}{{\ttfamily arXiv:2006.02445
  [quant-ph]}}.

\bibitem{Stuttard:2020qfv}
T.~Stuttard and M.~Jensen, ``{Neutrino decoherence from quantum gravitational
  stochastic perturbations},''
  \href{http://dx.doi.org/10.1103/PhysRevD.102.115003}{{\em Phys. Rev. D}
  {\bfseries 102} no.~11, (2020) 115003},
  \href{http://arxiv.org/abs/2007.00068}{{\ttfamily arXiv:2007.00068
  [hep-ph]}}.

\bibitem{Stuttard:2021uyw}
T.~Stuttard, ``{Neutrino signals of lightcone fluctuations resulting from
  fluctuating spacetime},''
  \href{http://dx.doi.org/10.1103/PhysRevD.104.056007}{{\em Phys. Rev. D}
  {\bfseries 104} no.~5, (2021) 056007},
  \href{http://arxiv.org/abs/2103.15313}{{\ttfamily arXiv:2103.15313
  [hep-ph]}}.

\bibitem{Banerjee:2022slh}
I.~K. Banerjee and U.~K. Dey, ``{Neutrino decoherence from generalised
  uncertainty},'' \href{http://dx.doi.org/10.1140/epjc/s10052-023-11565-0}{{\em
  Eur. Phys. J. C} {\bfseries 83} no.~5, (2023) 428},
  \href{http://arxiv.org/abs/2208.12062}{{\ttfamily arXiv:2208.12062
  [hep-ph]}}.

\bibitem{Hawking:1976ra}
S.~W. Hawking, ``{Breakdown of Predictability in Gravitational Collapse},''
  \href{http://dx.doi.org/10.1103/PhysRevD.14.2460}{{\em Phys. Rev. D}
  {\bfseries 14} (1976) 2460--2473}.

\bibitem{Ellis:1983jz}
J.~R. Ellis, J.~S. Hagelin, D.~V. Nanopoulos, and M.~Srednicki, ``{Search for
  Violations of Quantum Mechanics},''
  \href{http://dx.doi.org/10.1016/0550-3213(84)90053-1}{{\em Nucl. Phys. B}
  {\bfseries 241} (1984) 381}.

\bibitem{Giddings:1988cx}
S.~B. Giddings and A.~Strominger, ``{Loss of Incoherence and Determination of
  Coupling Constants in Quantum Gravity},''
  \href{http://dx.doi.org/10.1016/0550-3213(88)90109-5}{{\em Nucl. Phys. B}
  {\bfseries 307} (1988) 854--866}.

\bibitem{Addazi:2021xuf}
A.~Addazi {\em et~al.}, ``{Quantum gravity phenomenology at the dawn of the
  multi-messenger era\textemdash{}A review},''
  \href{http://dx.doi.org/10.1016/j.ppnp.2022.103948}{{\em Prog. Part. Nucl.
  Phys.} {\bfseries 125} (2022) 103948},
  \href{http://arxiv.org/abs/2111.05659}{{\ttfamily arXiv:2111.05659
  [hep-ph]}}.

\bibitem{deHolanda:2019tuf}
P.~C. de~Holanda, ``{Solar Neutrino Limits on Decoherence},''
  \href{http://dx.doi.org/10.1088/1475-7516/2020/03/012}{{\em JCAP} {\bfseries
  03} (2020) 012}, \href{http://arxiv.org/abs/1909.09504}{{\ttfamily
  arXiv:1909.09504 [hep-ph]}}.

\bibitem{deOliveira:2013dia}
R.~L.~N. de~Oliveira, M.~M. Guzzo, and P.~C. de~Holanda, ``{Quantum Dissipation
  and $C\!P$ Violation in MINOS},''
  \href{http://dx.doi.org/10.1103/PhysRevD.89.053002}{{\em Phys. Rev. D}
  {\bfseries 89} no.~5, (2014) 053002},
  \href{http://arxiv.org/abs/1401.0033}{{\ttfamily arXiv:1401.0033 [hep-ph]}}.

\bibitem{BalieiroGomes:2018gtd}
G.~Balieiro~Gomes, D.~V. Forero, M.~M. Guzzo, P.~C. De~Holanda, and R.~L.~N.
  Oliveira, ``{Quantum Decoherence Effects in Neutrino Oscillations at DUNE},''
  \href{http://dx.doi.org/10.1103/PhysRevD.100.055023}{{\em Phys. Rev. D}
  {\bfseries 100} no.~5, (2019) 055023},
  \href{http://arxiv.org/abs/1805.09818}{{\ttfamily arXiv:1805.09818
  [hep-ph]}}.

\bibitem{JUNO:2021ydg}
{\bfseries JUNO} Collaboration, J.~Wang {\em et~al.}, ``{Damping signatures at
  JUNO, a medium-baseline reactor neutrino oscillation experiment},''
  \href{http://dx.doi.org/10.1007/JHEP06(2022)062}{{\em JHEP} {\bfseries 06}
  (2022) 062}, \href{http://arxiv.org/abs/2112.14450}{{\ttfamily
  arXiv:2112.14450 [hep-ex]}}.

\bibitem{jonghee_yoo_2020_4123573}
J.~Yoo, ``Reno,'' June, 2020.
\newblock \url{https://doi.org/10.5281/zenodo.4123573}.

\bibitem{DayaBay:2018yms}
{\bfseries Daya Bay} Collaboration, D.~Adey {\em et~al.}, ``{Measurement of the
  Electron Antineutrino Oscillation with 1958 Days of Operation at Daya Bay},''
  \href{http://dx.doi.org/10.1103/PhysRevLett.121.241805}{{\em Phys. Rev.
  Lett.} {\bfseries 121} no.~24, (2018) 241805},
  \href{http://arxiv.org/abs/1809.02261}{{\ttfamily arXiv:1809.02261
  [hep-ex]}}.

\bibitem{Gando:2010aa}
{\bfseries KamLAND} Collaboration, A.~Gando {\em et~al.}, ``{Constraints on
  $\theta_{13}$ from A Three-Flavor Oscillation Analysis of Reactor
  Antineutrinos at KamLAND},''
  \href{http://dx.doi.org/10.1103/PhysRevD.83.052002}{{\em Phys. Rev. D}
  {\bfseries 83} (2011) 052002},
  \href{http://arxiv.org/abs/1009.4771}{{\ttfamily arXiv:1009.4771 [hep-ex]}}.

\bibitem{kamland_web}
``Kamland,'' 2010.
\newblock
  \url{https://www.awa.tohoku.ac.jp/KamLAND/4th_result_data_release/4th_result_data_release.html}.

\bibitem{Adamson:2017uda}
{\bfseries MINOS+} Collaboration, P.~Adamson {\em et~al.}, ``{Search for
  sterile neutrinos in MINOS and MINOS+ using a two-detector fit},''
  \href{http://dx.doi.org/10.1103/PhysRevLett.122.091803}{{\em Phys. Rev.
  Lett.} {\bfseries 122} no.~9, (2019) 091803},
  \href{http://arxiv.org/abs/1710.06488}{{\ttfamily arXiv:1710.06488
  [hep-ex]}}.

\bibitem{NOvA:2018gge}
{\bfseries NOvA} Collaboration, M.~Acero {\em et~al.}, ``{New constraints on
  oscillation parameters from $\nu_e$ appearance and $\nu_\mu$ disappearance in
  the NOvA experiment},''
  \href{http://dx.doi.org/10.1103/PhysRevD.98.032012}{{\em Phys.Rev. D}
  {\bfseries 98} (2018) 032012},
  \href{http://arxiv.org/abs/1806.00096}{{\ttfamily arXiv:1806.00096
  [hep-ex]}}.

\bibitem{NOvA:2021nfi}
{\bfseries NOvA} Collaboration, M.~A. Acero {\em et~al.}, ``{Improved
  measurement of neutrino oscillation parameters by the NOvA experiment},''
  \href{http://dx.doi.org/10.1103/PhysRevD.106.032004}{{\em Phys. Rev. D}
  {\bfseries 106} no.~3, (2022) 032004},
  \href{http://arxiv.org/abs/2108.08219}{{\ttfamily arXiv:2108.08219
  [hep-ex]}}.

\bibitem{Abe:2021gky}
{\bfseries T2K} Collaboration, K.~Abe {\em et~al.}, ``{Improved constraints on
  neutrino mixing from the T2K experiment with $\mathbf{3.13\times10^{21}}$
  protons on target},''
  \href{http://dx.doi.org/10.1103/PhysRevD.103.112008}{{\em Phys. Rev. D}
  {\bfseries 103} no.~11, (2021) 112008},
  \href{http://arxiv.org/abs/2101.03779}{{\ttfamily arXiv:2101.03779
  [hep-ex]}}.

\bibitem{JUNO:2015zny}
{\bfseries JUNO} Collaboration, F.~An {\em et~al.}, ``{Neutrino Physics with
  JUNO},'' \href{http://dx.doi.org/10.1088/0954-3899/43/3/030401}{{\em J. Phys.
  G} {\bfseries 43} no.~3, (2016) 030401},
  \href{http://arxiv.org/abs/1507.05613}{{\ttfamily arXiv:1507.05613
  [physics.ins-det]}}.

\bibitem{JUNO:2021vlw}
{\bfseries JUNO} Collaboration, A.~Abusleme {\em et~al.}, ``{JUNO physics and
  detector},'' \href{http://dx.doi.org/10.1016/j.ppnp.2021.103927}{{\em Prog.
  Part. Nucl. Phys.} {\bfseries 123} (2022) 103927},
  \href{http://arxiv.org/abs/2104.02565}{{\ttfamily arXiv:2104.02565
  [hep-ex]}}.

\bibitem{DUNE:2020lwj}
{\bfseries DUNE} Collaboration, B.~Abi {\em et~al.}, ``{Deep Underground
  Neutrino Experiment (DUNE), Far Detector Technical Design Report, Volume I
  Introduction to DUNE},''
  \href{http://dx.doi.org/10.1088/1748-0221/15/08/T08008}{{\em JINST}
  {\bfseries 15} no.~08, (2020) T08008},
  \href{http://arxiv.org/abs/2002.02967}{{\ttfamily arXiv:2002.02967
  [physics.ins-det]}}.

\bibitem{DUNE:2020ypp}
{\bfseries DUNE} Collaboration, B.~Abi {\em et~al.}, ``{Deep Underground
  Neutrino Experiment (DUNE), Far Detector Technical Design Report, Volume II:
  DUNE Physics},'' \href{http://arxiv.org/abs/2002.03005}{{\ttfamily
  arXiv:2002.03005 [hep-ex]}}.

\bibitem{DUNE:2020mra}
{\bfseries DUNE} Collaboration, B.~Abi {\em et~al.}, ``{Deep Underground
  Neutrino Experiment (DUNE), Far Detector Technical Design Report, Volume III:
  DUNE Far Detector Technical Coordination},''
  \href{http://dx.doi.org/10.1088/1748-0221/15/08/T08009}{{\em JINST}
  {\bfseries 15} no.~08, (2020) T08009},
  \href{http://arxiv.org/abs/2002.03008}{{\ttfamily arXiv:2002.03008
  [physics.ins-det]}}.

\bibitem{DUNE:2020txw}
{\bfseries DUNE} Collaboration, B.~Abi {\em et~al.}, ``{Deep Underground
  Neutrino Experiment (DUNE), Far Detector Technical Design Report, Volume IV:
  Far Detector Single-phase Technology},''
  \href{http://dx.doi.org/10.1088/1748-0221/15/08/T08010}{{\em JINST}
  {\bfseries 15} no.~08, (2020) T08010},
  \href{http://arxiv.org/abs/2002.03010}{{\ttfamily arXiv:2002.03010
  [physics.ins-det]}}.

\bibitem{Alicki:1105909}
R.~Alicki and K.~Lendi, \href{http://dx.doi.org/10.1007/3-540-70861-8}{{\em
  {Quantum dynamical semigroups and applications}}}.
\newblock Lecture notes in physics. Springer, Berlin, 2007.
\newblock \url{https://cds.cern.ch/record/1105909}.

\bibitem{Lindblad:1975ef}
G.~Lindblad, ``{On the Generators of Quantum Dynamical Semigroups},''
  \href{http://dx.doi.org/10.1007/BF01608499}{{\em Commun. Math. Phys.}
  {\bfseries 48} (1976) 119}.

\bibitem{Gorini:1975nb}
V.~Gorini, A.~Kossakowski, and E.~C.~G. Sudarshan, ``{Completely Positive
  Dynamical Semigroups of N Level Systems},''
  \href{http://dx.doi.org/10.1063/1.522979}{{\em J. Math. Phys.} {\bfseries 17}
  (1976) 821}.

\bibitem{Gago:2002na}
A.~M. Gago, E.~M. Santos, W.~J.~C. Teves, and R.~Zukanovich~Funchal, ``{A Study
  on quantum decoherence phenomena with three generations of neutrinos},''
  \href{http://arxiv.org/abs/hep-ph/0208166}{{\ttfamily arXiv:hep-ph/0208166}}.

\bibitem{Barenboim:2004wu}
G.~Barenboim and N.~E. Mavromatos, ``{CPT violating decoherence and LSND: A
  Possible window to Planck scale physics},''
  \href{http://dx.doi.org/10.1088/1126-6708/2005/01/034}{{\em JHEP} {\bfseries
  01} (2005) 034}, \href{http://arxiv.org/abs/hep-ph/0404014}{{\ttfamily
  arXiv:hep-ph/0404014}}.

\bibitem{Barenboim:2006xt}
G.~Barenboim, N.~E. Mavromatos, S.~Sarkar, and A.~Waldron-Lauda, ``{Quantum
  decoherence and neutrino data},''
  \href{http://dx.doi.org/10.1016/j.nuclphysb.2006.09.012}{{\em Nucl. Phys. B}
  {\bfseries 758} (2006) 90--111},
  \href{http://arxiv.org/abs/hep-ph/0603028}{{\ttfamily arXiv:hep-ph/0603028}}.

\bibitem{Mavromatos:2006yn}
N.~E. Mavromatos and S.~Sarkar, ``{Probing Models of Quantum Decoherence in
  Particle Physics and Cosmology},''
\newblock 12, 2006.
\newblock \href{http://arxiv.org/abs/hep-ph/0612193}{{\ttfamily
  arXiv:hep-ph/0612193}}.

\bibitem{Oliveira:2010zzd}
R.~L.~N. Oliveira and M.~M. Guzzo, ``{Quantum dissipation in vacuum neutrino
  oscillation},'' \href{http://dx.doi.org/10.1140/epjc/s10052-010-1388-1}{{\em
  Eur. Phys. J. C} {\bfseries 69} (2010) 493--502}.

\bibitem{Benatti:1987dz}
F.~Benatti and H.~Narnhofer, ``{ENTROPY BEHAVIOR UNDER COMPLETELY POSITIVE
  MAPS},'' \href{http://dx.doi.org/10.1007/BF00419590}{{\em Lett. Math. Phys.}
  {\bfseries 15} (1988) 325}.

\bibitem{Guzzo:2014jbp}
M.~M. Guzzo, P.~C. de~Holanda, and R.~L.~N. Oliveira, ``{Quantum Dissipation in
  a Neutrino System Propagating in Vacuum and in Matter},''
  \href{http://dx.doi.org/10.1016/j.nuclphysb.2016.04.030}{{\em Nucl. Phys. B}
  {\bfseries 908} (2016) 408--422},
  \href{http://arxiv.org/abs/1408.0823}{{\ttfamily arXiv:1408.0823 [hep-ph]}}.

\bibitem{Liu:1997km}
Y.~Liu, L.-z. Hu, and M.-L. Ge, ``{The Effect of quantum mechanics violation on
  neutrino oscillation},''
  \href{http://dx.doi.org/10.1103/PhysRevD.56.6648}{{\em Phys. Rev. D}
  {\bfseries 56} (1997) 6648--6652}.

\bibitem{Ellis:1995xd}
J.~R. Ellis, J.~L. Lopez, N.~E. Mavromatos, and D.~V. Nanopoulos, ``{Precision
  tests of CPT symmetry and quantum mechanics in the neutral kaon system},''
  \href{http://dx.doi.org/10.1103/PhysRevD.53.3846}{{\em Phys. Rev. D}
  {\bfseries 53} (1996) 3846--3870},
  \href{http://arxiv.org/abs/hep-ph/9505340}{{\ttfamily arXiv:hep-ph/9505340}}.

\bibitem{Ellis:2000dy}
J.~R. Ellis, N.~E. Mavromatos, and D.~V. Nanopoulos, ``{How large are
  dissipative effects in noncritical Liouville string theory?},''
  \href{http://dx.doi.org/10.1103/PhysRevD.63.024024}{{\em Phys. Rev. D}
  {\bfseries 63} (2001) 024024},
  \href{http://arxiv.org/abs/gr-qc/0007044}{{\ttfamily arXiv:gr-qc/0007044}}.

\bibitem{Gambini:2003pv}
R.~Gambini, R.~A. Porto, and J.~Pullin, ``{Decoherence from discrete quantum
  gravity},'' \href{http://dx.doi.org/10.1088/0264-9381/21/8/L01}{{\em Class.
  Quant. Grav.} {\bfseries 21} (2004) L51--L57},
  \href{http://arxiv.org/abs/gr-qc/0305098}{{\ttfamily arXiv:gr-qc/0305098}}.

\bibitem{Ellis:1996bz}
J.~R. Ellis, N.~E. Mavromatos, D.~V. Nanopoulos, and E.~Winstanley, ``{Quantum
  decoherence in a four-dimensional black hole background},''
  \href{http://dx.doi.org/10.1142/S0217732397000248}{{\em Mod. Phys. Lett. A}
  {\bfseries 12} (1997) 243--256},
  \href{http://arxiv.org/abs/gr-qc/9602011}{{\ttfamily arXiv:gr-qc/9602011}}.

\bibitem{Ellis:1997jw}
J.~R. Ellis, N.~E. Mavromatos, and D.~V. Nanopoulos, ``{Quantum decoherence in
  a D foam background},''
  \href{http://dx.doi.org/10.1142/S0217732397001795}{{\em Mod. Phys. Lett. A}
  {\bfseries 12} (1997) 1759--1773},
  \href{http://arxiv.org/abs/hep-th/9704169}{{\ttfamily arXiv:hep-th/9704169}}.

\bibitem{Farzan:2023fqa}
Y.~Farzan and T.~Schwetz, ``{A decoherence explanation of the gallium neutrino
  anomaly},'' \href{http://arxiv.org/abs/2306.09422}{{\ttfamily
  arXiv:2306.09422 [hep-ph]}}.

\bibitem{Giunti:2022btk}
C.~Giunti, Y.~F. Li, C.~A. Ternes, O.~Tyagi, and Z.~Xin, ``{Gallium Anomaly:
  critical view from the global picture of \ensuremath{\nu}$_{e}$ and $
  {\overline{\nu}}_e $ disappearance},''
  \href{http://dx.doi.org/10.1007/JHEP10(2022)164}{{\em JHEP} {\bfseries 10}
  (2022) 164}, \href{http://arxiv.org/abs/2209.00916}{{\ttfamily
  arXiv:2209.00916 [hep-ph]}}.

\bibitem{Denton:2016wmg}
P.~B. Denton, H.~Minakata, and S.~J. Parke, ``{Compact Perturbative Expressions
  For Neutrino Oscillations in Matter},''
  \href{http://dx.doi.org/10.1007/JHEP06(2016)051}{{\em JHEP} {\bfseries 06}
  (2016) 051}, \href{http://arxiv.org/abs/1604.08167}{{\ttfamily
  arXiv:1604.08167 [hep-ph]}}.

\bibitem{Barenboim:2019pfp}
G.~Barenboim, P.~B. Denton, S.~J. Parke, and C.~A. Ternes, ``{Neutrino
  Oscillation Probabilities through the Looking Glass},''
  \href{http://dx.doi.org/10.1016/j.physletb.2019.03.002}{{\em Phys. Lett. B}
  {\bfseries 791} (2019) 351--360},
  \href{http://arxiv.org/abs/1902.00517}{{\ttfamily arXiv:1902.00517
  [hep-ph]}}.

\bibitem{Huber:2004ka}
P.~Huber, M.~Lindner, and W.~Winter, ``{Simulation of long-baseline neutrino
  oscillation experiments with GLoBES (General Long Baseline Experiment
  Simulator)},'' \href{http://dx.doi.org/10.1016/j.cpc.2005.01.003}{{\em
  Comput. Phys. Commun.} {\bfseries 167} (2005) 195},
\href{http://arxiv.org/abs/hep-ph/0407333}{{\ttfamily arXiv:hep-ph/0407333
  [hep-ph]}}.

\bibitem{Huber:2007ji}
P.~Huber, J.~Kopp, M.~Lindner, M.~Rolinec, and W.~Winter, ``{New features in
  the simulation of neutrino oscillation experiments with GLoBES 3.0: General
  Long Baseline Experiment Simulator},''
  \href{http://dx.doi.org/10.1016/j.cpc.2007.05.004}{{\em Comput. Phys.
  Commun.} {\bfseries 177} (2007) 432--438},
\href{http://arxiv.org/abs/hep-ph/0701187}{{\ttfamily arXiv:hep-ph/0701187
  [hep-ph]}}.

\bibitem{Huber:2004xh}
P.~Huber and T.~Schwetz, ``{Precision spectroscopy with reactor
  anti-neutrinos},'' \href{http://dx.doi.org/10.1103/PhysRevD.70.053011}{{\em
  Phys. Rev.} {\bfseries D70} (2004) 053011},
\href{http://arxiv.org/abs/hep-ph/0407026}{{\ttfamily arXiv:hep-ph/0407026
  [hep-ph]}}.

\bibitem{Vogel:1999zy}
P.~Vogel and J.~F. Beacom, ``The angular distribution of the neutron inverse
  beta decay, $\overline{\nu}_e + p \to e^+ + n$,'' {\em Phys. Rev.} {\bfseries
  D60} (1999) 053003,
\href{http://arxiv.org/abs/hep-ph/9903554}{{\ttfamily hep-ph/9903554}}.

\bibitem{IceCube-Gen2:2019fet}
{\bfseries IceCube-Gen2} Collaboration, M.~G. Aartsen {\em et~al.}, ``{Combined
  sensitivity to the neutrino mass ordering with JUNO, the IceCube Upgrade, and
  PINGU},'' \href{http://dx.doi.org/10.1103/PhysRevD.101.032006}{{\em Phys.
  Rev. D} {\bfseries 101} no.~3, (2020) 032006},
  \href{http://arxiv.org/abs/1911.06745}{{\ttfamily arXiv:1911.06745
  [hep-ex]}}.

\bibitem{Forero:2021lax}
D.~V. Forero, S.~J. Parke, C.~A. Ternes, and R.~Z. Funchal,
  ``{JUNO\textquoteright{}s prospects for determining the neutrino mass
  ordering},'' \href{http://dx.doi.org/10.1103/PhysRevD.104.113004}{{\em Phys.
  Rev. D} {\bfseries 104} no.~11, (2021) 113004},
  \href{http://arxiv.org/abs/2107.12410}{{\ttfamily arXiv:2107.12410
  [hep-ph]}}.

\bibitem{DUNE:2021cuw}
{\bfseries DUNE} Collaboration, B.~Abi {\em et~al.}, ``{Experiment Simulation
  Configurations Approximating DUNE TDR},''
  \href{http://arxiv.org/abs/2103.04797}{{\ttfamily arXiv:2103.04797
  [hep-ex]}}.

\bibitem{DUNEHE}
``{DUNE Tau-Optimized Fluxes}.''
\newblock \url{https://glaucus.crc.nd.edu/DUNEFluxes/TauOptimized/}.

\end{thebibliography}

\providecommand{\href}[2]{#2}\begingroup\raggedright\endgroup

\end{document}